\documentclass[fleqn,usenatbib,useAMS]{mnras}
\usepackage{graphicx}	
\usepackage{amsmath}	
\usepackage{amssymb}	
\usepackage{multicol}        
\usepackage{bm}		
\usepackage{pdflscape}	
\usepackage[T1]{fontenc}
\usepackage{ae,aecompl}
\usepackage{newtxtext,newtxmath}
\usepackage{multirow}
\usepackage{numprint}

\title[Effect of stellar spots on ARIEL observations]{Correcting the effect of stellar spots on ARIEL transmission spectra}

\author[G. Cracchiolo et al.]{
G. Cracchiolo,$^{1,2}$\thanks{E-mail: gianluca.cracchiolo@inaf.it}
G. Micela,$^{2}$
G. Peres$^{1}$
\\
$^{1}$Dipartimento di Fisica e Chimica, Università degli studi di Palermo, Via Archirafi 36, 90123 Palermo, Italy\\
$^{2}$INAF-Osservatorio Astronomico di Palermo, Piazza del Parlamento 1, 90134 Palermo, Italy
}

\date{Accepted 2020 November 17. Received 2020 November 16; in original form 2020 July 6}

\pubyear{2020}
\begin{document}
\label{firstpage}
\pagerange{\pageref{firstpage}--\pageref{lastpage}}
\maketitle

\begin{abstract}
    The goal of this study is to assess the impact of the stellar spots on the extraction of the planetary transmission spectra observed by ARIEL. We develop a method to model the stellar spectrum of a star in the presence of spots by using the out-of-transit observations. It is based on a chi squared minimization procedure of the out-of-transit spectrum on a grid of stellar spectra with different sizes and temperatures of the spots. The approach allows us also to study the temporal evolution of the spots when comparing stellar spectra observed at different epochs. We also present a method to correct the transit depth variations due to non-occulted stellar spots and estimate the error we introduce if we apply the same correction to crossings over the stellar spots. The method is tested on three types of stellar targets that ARIEL will observe in its 4-years mission lifetime. In all the explored cases, the approach allows to reliably recover the spot parameters (size and temperature) from out-of-transit observations and, for non-occulted spots, to confidently recover the planetary atmosphere transmission spectrum within the noise level(with average uncertainty of at most $3.3\%$ of the planetary signal). Conversely, we find systematic biases in the inferred planetary spectra due to the occulted spots, with measurable effects for the brightest targets especially for more contrasted spots.
\end{abstract}


\section{Introduction}\label{Introduction}
The ARIEL mission (Atmospheric Remote-sensing Infrared
Exoplanet Large-survey, \citealt{ARIEL}) has been selected
as the fourth Medium Class Mission of the Cosmic Vision program of the European Space Agency (ESA). During its 4-years mission, ARIEL will survey a diverse sample of about $1\,000$ known extrasolar planets ranging from Jupiters and Neptunes down to super-Earth size orbiting different types of stars, by using simultaneously three photometric bands in the visible ($0.50 - 1.10 \;\mu m$) and three spectroscopic channels in the infrared
($1.10-7.80 \;\mu m$). It is the first mission dedicated to measuring the chemical composition, thermal structures and scattering properties of the atmospheres of these  planets. ARIEL will enable the study of exoplanets through transit, eclipse, and phase-curve observations (see, e.g., \citealt{description_observation} for a complete description of these techniques).\\

Here, we focus on ARIEL’s transits observations and we analyze the impact of stellar activity on the transit depth deriving from spots, i.e. cooler and darker features on the stellar surface, caused by stellar magnetic activity and with a lower temperature than the surrounding photosphere. Both occulted and non-occulted spots can bias the measured transit depth in opposite directions: non-occulted spots will cause the transit depth to be overestimated, while occulted spots will appear in the transit light curve as an upward “bump” lasting roughly as long as it takes the planet to cross the spot, and may lead to underestimate the planetary radius\footnote{The opposite effect will occur in case of regions hotter than the photosphere, e.g., faculae.} (e.g., see \citealt{Pont} for a more detailed description). The bias due to both occulted and non-occulted spots on the transit depth is chromatic, with stronger effects towards shorter wavelengths, and it may have a significant impact on our knowledge of the dependence of the planetary radius on the wavelength, thus hampering the extraction of the final transmission spectrum of the planetary atmosphere (\citealt{Algol, Berta, Czesla, Desert, Pont, Sing_2009}). So the issue is to understand if the apparent radius variations in an observation are due to the presence of molecular species in the planetary atmospheres or to stellar spots whose effects are also expected to depend on wavelength. The existence of activity-related “pseudo”-transmission effects is well known and has been discussed, e.g., in \cite{Salz_2018}, in the context of high-resolution atomic lines. They show that stellar activity related pseudo-signals mix with and confuse the planetary atmospheric absorption signal. In some cases, pseudo-effects can be responsible for the major fraction of the signal. In this work we discuss a technique to correct the variations of the transit depth due to  non-occulted spots, based on correction factors obtained modeling the spectrum of the spot and the photosphere (see Section~\ref{method}). We won't present any approach to correct the transits over the spots, but we will estimate the errors we introduce if we correct the transits over the spots with the same technique used for non-occulted spots. The approach is purposely developed and tested on ARIEL-like spectra, generated with the Ariel radiometric model (ArielRad, \citealt{ArielRad}), a software simulating the stellar signal in the ARIEL photometric and spectral channels and that implements a realistic noise model.\\

A further problem, in the presence of spots, one has to take care of is the variability of the stellar flux caused by the stellar rotation or by the evolution of spots, on timescale of weeks/months, or by the stellar magnetic cycles, on a timescale of years. Due to these phenomena, the observed stellar flux may change according to the spotted area on the stellar hemisphere exposed to the observer. Almost all the existing methods used to filter out these kinds of long-term stellar flux variability are based on the observations of the flux variations outside transits, along time intervals of at least two or three stellar rotations (e.g., \citealt{Aigrain, Bonomo_2008, Bonomo_2009, Moutou}). Typically the flux is monitored in the visible range, where variations induced by the stellar spots are stronger than in the infrared. The variations in flux caused by this effect are typically of the order of a few percent in active stars in the visible range. For example, for the active dK HD 189733, \cite{Pont} noted a $1-2\;\%$ flux reduction in the visible range due to spots along its $\sim 12$ days rotation cycle, which is comparable to the planet transit depth. However, in some cases, the effects of spot-related variability were assessed to be not significant, as for the case GJ 1214 b (\citealt{Kreidberg_2014}) which showed consistent transit depths over all epochs. Here we will try to account also for the temporal evolution of the spots from the comparison between out-of-transit observations taken at different epochs.\\

In Section~\ref{method} we present our model of stellar photospheric activity. In Section~\ref{sim_out_transit} we describe the algorithm to derive the spot parameters. In Section~\ref{retrieval_planet} we show the capability of the method to correct the transit depth variations due to non-occulted spots and we discuss the limits of applicability of our method to spot crossing events. Finally, Section~\ref{Summary} contains a summary of the obtained results.

\section{Modeling the stellar spectrum}\label{method}
Our aim is to develop a method to derive the stellar spots physical and geometrical properties from out-of-transit observations, possibly taking into account the temporal evolution of the spots between observations of the same star at different epochs. This information will be used to correctly derive the planetary spectrum  obtained during the transit.

\subsection{Active star models}
The modeling of the stellar activity is based on the assumption that it is dominated by the presence of spots at temperature $T_s$ covering a fraction $ff$ (filling factor) of the visible stellar disk. The out-of transit stellar flux $F_\lambda^{out}$ in the presence of spots can be expressed as in \cite{Ballerini}, \cite{Micela}:
\begin{equation}
    F_\lambda^{out}=(1-ff)\cdot F_\lambda(T_*)+ ff\cdot F_\lambda(T_s)
    \label{model_out}
\end{equation}
where $T_*$ is the effective stellar temperature for a given spectral type star in the absence of spots, $T_s$ is the spot temperature (with $T_s<T_*$) and $ff$ is the filling factor, i.e. the fraction of the projected stellar surface covered by the spots with $0 \leq ff \leq 1$. Eq.~(\ref{model_out}) treats each of the two components (photosphere and spots) as having uniform surface brightness and it neglects the effects of brighter active regions on the star, faculae and plages, i.e. photospheric inhomogeneities with temperatures higher than the unperturbed photosphere, or the effect of the limb darkening.\\
Due to the stellar rotation and changes in the stellar activity, we may have changes in the filling factor $ff$ between different observations on the timescale of the stellar rotation period and of the spots evolution ($\sim$ weeks/months). 
The timescale for a rotating spot going from full to zero projected area is 1/4 the stellar rotation period. For most stars, this remains generally longer than the typical transit timescale (a few hours), so the filling factor variations on the transit timescale may be neglected. This implies that the spot pattern does not change during the transit and the measured stellar flux variations can be attributable only to the planet occultation and can be considered constant during the transit\footnote{However, it should be stressed that observations of highly active stars, e.g. CoRoT-2 (\citealt{CoRoT-2}), have shown that significant rotational variations can take place on the transit timescale, so for these very active targets our assumption cannot be made.}. On the contrary, the spot configuration can change between different transits.

\subsection{Active star model during primary transit}\label{primary_transit}
In general, during the primary transit in the presence of a spot on the stellar surface, the planet can occult part of the radiation coming from the spot and part coming from the unperturbed photosphere. Since the planetary radius is wavelength-dependent, the fraction $g_\lambda$ of the planet that occults the spot at time $t$ is wavelength-dependent (with $0\leq g_\lambda\leq 1$). So, the measured flux of the system “spotted-star + planet” $F_{\lambda}^{in}$ can be expressed as:
\begin{equation}
\label{model_in_transit}
F_{\lambda}^{in}   =                                                    [1-ff-(1-g_{\lambda})\cdot\varepsilon_{\lambda}] \cdot            F_{\lambda}(T_{*}) + (ff                                          -g_{\lambda}\cdot\varepsilon_{\lambda})\cdot                                    F_{\lambda}(T_{s})
\end{equation}
where $\varepsilon_{\lambda}=\left(\frac{\pi R_{p}^{2}}{\pi R_{*}^{2}}\right)_{\lambda}$ is the ratio between the areas of the planetary and of the stellar disks, respectively, at each wavelength and is important in that it contains the information about the size of the planet and of its atmosphere.\\
In the following, we will assume that the stellar spots physical and geometrical properties out-of-transit are not significantly different from those observed during the transit. This is a reasonable hypothesis as discussed in the previous Section. Under this assumption we can express the in-transit spectrum as follow:
\begin{equation}
\label{model_in_transit2}
    F_{\lambda}^{in}   = F_{\lambda}^{out}-\varepsilon_{\lambda}\cdot[(1-g_{\lambda}) \cdot F_{\lambda}(T_{*}) + g_{\lambda}\cdot F_{\lambda}(T_{s})]
\end{equation}
The planetary signal $\varepsilon_{\lambda}$ can be derived from the previous equation:
\begin{equation}
\label{planet}
    \begin{split}
    \varepsilon_{\lambda}  &=\frac{F_{\lambda}^{out}-F_{\lambda}^{in}}{(1-g_{\lambda}) \cdot F_{\lambda}(T_{*})+g_{\lambda}\cdot F_{\lambda}(T_{s})}\\
    &= \delta_\lambda \cdot \frac{(1-ff)\cdot F_{\lambda}(T_{*})+ff\cdot F_{\lambda}(T_{s})}{(1-g_{\lambda}) \cdot F_{\lambda}(T_{*})+g_{\lambda}\cdot F_{\lambda}(T_{s})}
    \end{split}
\end{equation}
where $\delta_\lambda=\frac{F_\lambda^{out}-F_\lambda^{in}}{F_\lambda^{out}}$ is the uncorrected transit depth, while the second term is the correction factor that gives the ratio of the flux of the spotted star to the flux of the star covered by the planet. Generally, $g_\lambda$ will change with time within the transit, as the planet transit chord passes over randomly distributed spots of different sizes. When the planet transits out of the spot, $g_{\lambda}=0$ at every time $t$, therefore the correction factor reduces to $\left[1+ff\cdot\left(\frac{F_\lambda(T_s)}{F_\lambda(T_*)}-1\right)\right]$. A similar approach can be found in \cite{McCullough, Pont, Zellem}. If the planet crosses the spot during the transit, $g_\lambda$ varies over time. In the latter case, knowing $g_\lambda$ at all instants would allow us to correct the transit depth in all the points of the light curve.

\subsection{Ariel Radiometric model}\label{arielrad}
The method is tested on three possible star-planet combinations, using as examples three transiting systems, potential ARIEL targets (see Table~\ref{input_param}), taken from \cite{Edwards} catalogue. For each star, a grid of stellar spectra out-of-transit is generated as in Eq.~(\ref{model_out}), where the spectrum $F_{\lambda}(T)$ of the unperturbed photosphere and the spots spectra are generated with the Ariel Radiometric Model (\citealt{ArielRad}). In this paper we define “spectrum” the set of flux values collected simultaneously in the photometric channels and in the low resolution spectral channels.\\

ArielRad selects the stellar model-atmosphere by using BT-Settl models (\url{https://phoenix.ens-lyon.fr/Grids/BT-Settl/}, \citealt{Baraffe}), where the stellar metallicity is $[M/H] =0.0$, corresponding to the solar abundances. The effective temperature in the BT-Settl grid spans the $1200\;K - 7000\;K$ interval, while the logaritm of the surface gravity $g$ ranges in the interval $[2.5 - 5.5]$. Spectra corresponding to temperatures not present in the Phoenix library are generated by interpolating spectra with the closest temperatures, while models corresponding to missing values of $\log g$ are selected by choosing the model with the closest value. By taking into account the distance $D$ of the star and the instrumental properties, ArielRad simulates the stellar flux $F_{\lambda}(T)$ (in $counts/sec$) in ARIEL's photometric and spectral channels (Table~\ref{Instr}).  The software simulates ARIEL Level 2 data, i.e. data that are corrected for instrumental effects (bad or saturated pixels, pointing jitter, etc.), and wavelength calibrated (see \citealt{Data_level}, submitted, for more details on ARIEL Data Levels).\\

\begin{table}
    \centering
    \caption{Charateristics of the instruments on ARIEL's focal plane: FGS (“Fine Guidance System”) and AIRS (“ARIEL IR Spectrometer”). The first three channels are the three ARIEL's photometers. The spectral resolution $R$ is shown in the last column.}
    \begin{tabular}{cccc}
    \hline
          Instrument          &Channel   &$\Delta \lambda$   & Resolution\\
    \hline
        \multirow{4}{4em}{FGS}  & VISPhot  & $0.50-0.60\;\mu m$  &  1  \\
                                & FGSPrime & $0.60-0.80\;\mu m$  &  1  \\
                                & FGSRed   & $0.80-1.10\;\mu m$  &  1  \\
                                & NIRSpec  & $1.10-1.95\;\mu m$  & 20  \\
        \cline{2-4}
        \multirow{2}{4em}{AIRS} & AIRSCh0  & $1.95-3.90\;\mu m$ & 100 \\
                                & AIRSCh1  & $3.90-7.80\;\mu m$   & 30  \\
    \hline
    \end{tabular}
    \label{Instr}
\end{table}

The spectrum $F_{\lambda}(T_{*})$ of the non-spotted photospheres is generated by setting the input parameters as shown in Table~\ref{input_param}. We assume that the spot spectrum is equivalent to a stellar spectrum with a lower temperature, therefore the fluxes $F_{\lambda}(T_{s})$ are generated as for $F_{\lambda}(T_{*})$ changing the input stellar temperature. For each star we build a 2D grid of spotted spectra in the $ff - \Delta T$ space, where $\Delta T=T_{*}-T_{s}$. In this space, the filling factor $ff$ of the spots spans the $[0.001; 0.200]$ interval with a step of $\delta ff = 0.001$, while the temperature $T_{s}$ ranges in the $T_{*}-2000\;K;\; T_{*}-50\;K$ interval with a step of $\delta T=50\;K$. The spots parameters are summarized in Table~\ref{input_param}. \\

The spectrum $F_{\lambda}(T_{*})$ of the non-spotted photospheres is generated by setting the input parameters as shown in Table~\ref{input_param}. We assume that the spot spectrum is equivalent to a stellar spectrum with a lower temperature, therefore the fluxes $F_{\lambda}(T_{s})$ are generated as for $F_{\lambda}(T_{*})$ changing the input stellar temperature. For each star we build a 2D grid of spotted spectra in the $ff - \Delta T$ space, where $\Delta T=T_{*}-T_{s}$. In this space, the filling factor $ff$ of the spots spans the $[0.001; 0.200]$ interval with a step of $\delta ff = 0.001$, while the temperature $T_{s}$ ranges in the $T_{*}-2000\;K;\; T_{*}-50\;K$ interval with a step of $\delta T=50\;K$. The spots parameters are summarized in Table~\ref{input_param}. \\

\begin{table}
    \centering
    \caption{$T_{*}$, $R_{*}$, $M_{*}$ and $D$ are the input parameters used to generate the fluxes $F_{\lambda}$ of the non-spotted photosphere for the three stars considered.
    $V$ is the apparent magnitude in the visible band. All the stellar parameters are taken from the \citet{Edwards} catalogue. $T_{s}$ and $ff$ summarize the spots parameters used to build
    the out-of-transit stellar spectra grids.}
    \begin{tabular}{cccc}
    \hline
                         & HD 17156         & HAT-P-11       & K2-21\\
    \hline
    Spectral type        & G0               & K4             & M0 \\
    $V$                  & 8.2              & 9.6            & 12.8 \\
    $T_{*}(K)$           &$6040$            &$4780$          & $4220$\\
    $R_{*}(R_{\odot})$   &$1.55$            &$0.68$          & $0.65$ \\
    $M_{*}(M_{\odot})$   &$1.41$            &$0.81$          & $0.68$\\
    $D(pc)$              &$75.0$            &$37.8$          & $83.9$\\
    $T_{s}(K)$           &$4040-5990$       &$2780-4730$     & $2220-4170$\\
    $ff$                 &$0.001-0.200$     &$0.001-0.200$   & $0.001-0.200$ \\
    \hline
    \end{tabular}
    \label{input_param}
\end{table}

In this work we use ArielRad version 1.5r14. The software simulator provides a comprehensive model of the instrument performance and implements a realistic noise model. The noise sources simulated by ArielRad are expressed as a relative noise over an integration time of one hour, so it has to be scaled over the exposure time $t_{exp}$ of the observation. We indicate all known noise sources generated by ArielRad as $kn$. Furthermore, there are other sources of noise which are not simulated by ArielRad and that costitute the ARIEL noise floor $nf$, i.e. the lowest possible signal level that ARIEL may measure. The relative noise $rn$ associated to an observation has to take into accounts both terms as follows (see Eq. (17) of \citealt{ArielRad}):
\begin{equation}
    \label{noise}
    rn=\sqrt{\frac{kn^{2}}{t_{exp}(hr)}+nf^{2}}
\end{equation}
where $t_{exp}(hr)$ is the exposure time of the observation expressed in hours. ARIEL noise floor is conservatively set to $20 \;ppm$.

\section{Simulations of Active Stars}\label{sim_out_transit}
In order to test the capability of our method to retrieve the spot parameters from an observed out-of-transit spectrum, in the first part of this Section we simulate a set of spotted spectra with different combinations of $ff$ and $T_{s}$ and realistic noise, generated with Eq.~(\ref{noise}). We fit each noisy spectrum by using the derivative-free Nelder-Mead method (or downhill simplex; \citealt{Nelder}). The minimization algorithm is implemented with the python package \texttt{scipy} (\citealt{scipy}) and it minimizes the $\chi^2(ff, T_s)$ function between each noisy spectrum and the bi-dimensional grid of models. We make each $\chi^2$ function continuous, by linearly interpolating the stellar spectra in the 2D grid, so the minimization method searches the solution over a continuous parameter space and it is not constrained on the grid nodes. The grid of spectra built as shown in Section~\ref{method} is thus a bi-dimensional collection of reference spectra for the fitting procedure.\\

The downhill simplex method does not handle boundary conditions, so the fitted spot parameters that minimize the $\chi^2$ function may escape off their grid. In order to prevent this, we impose that each $\chi^2$ function is infinite outside the boundaries of the 2D grid so that the solution is always within the spot parameters domain. The fitting algorithm has been employed in other works (e.g., \citealt{Scandariato2015, Example}) and similar fitting approaches can be found in \cite{O_Neal_1998}, where the star spotted area and temperature are determined by modeling some specific molecular absorption bands.\\

This technique allows us to determine the spectrum changes due to the presence of spots and it may provide information about the temporal evolution of the spots when comparing different observations. Using this info in the  next Section, we will simulate noisy transit observations as in Eq.~(\ref{model_in_transit}), and we will show the method capability in retrieving the planetary signal in the presence of non-occulted spots.

\subsection{Retrieval of the spot parameters} 
We evaluate the capabilities of our method for different spectral types, by fixing the signal-to-noise (S/N) ratio, i.e. we assume a visible magnitude $V=9$ and an exposure time $t_{exp}=100\;sec$. For each combination $T_{s}$, $ff$ we generate $10\,000$ random realizations of ARIEL observed fluxes taking into account the noise to be assigned to the simulated spectrum, and we run our fitting algorithm on each noisy spectrum. The spectra in the grids and the simulated spectra are normalized to their integrated fluxes.\\

\begin{figure*}
	{\includegraphics[width=\textwidth]{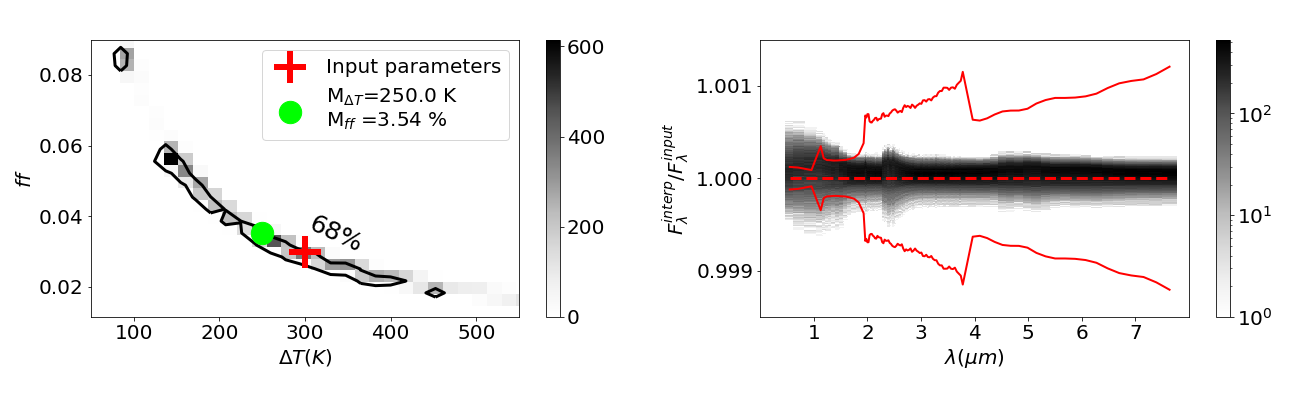}\quad
	\includegraphics[width=\textwidth]{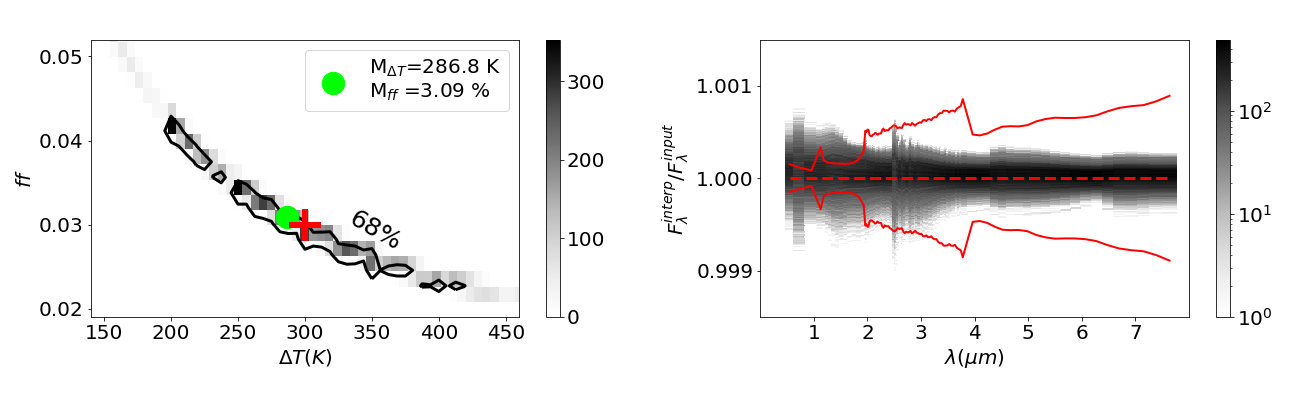}\quad
	\includegraphics[width=\textwidth]{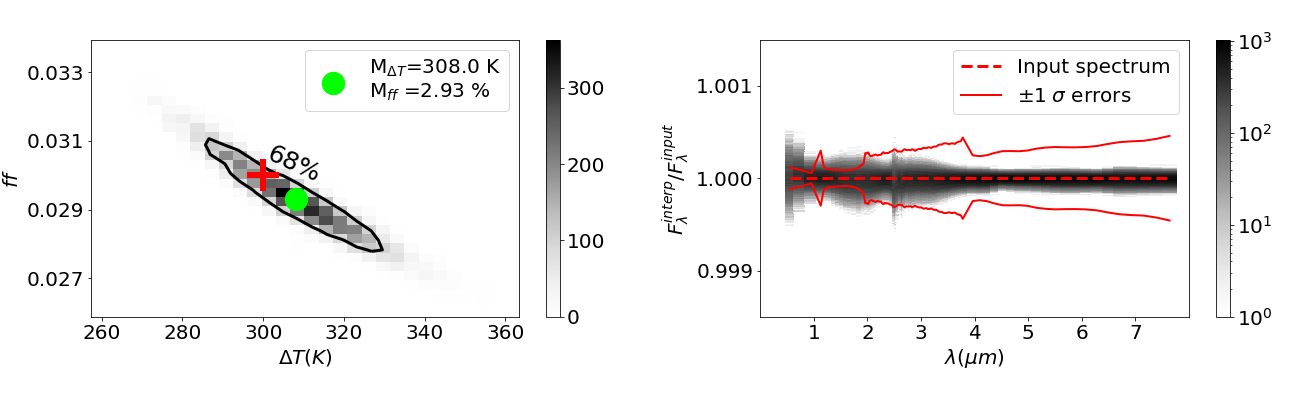}}
    \caption{Statistical analysis of the results from our fitting algorithm run over $10\,000$ random realizations of noisy spectra  with $100 sec$ exposures, in stars with $V=9$. The simulated configuration has a spot temperature $300\;K$ cooler than the photosphere and $ff = 0.03$, for the HD 17156 star (top), HAT-P-11 star (center) and K2-21 star (bottom), respectively. \textit{Left panels} - 2D probability distribution functions of the fitted parameters. The red cross represents the input parameters, while the green point marks the median of the distribution. The black contour shows the confidence region around the most probable value of the 2D distribution containing the 68\% of the sample. \textit{Right panels} - The 2D distribution shows the ratio between the best-fit spectra obtained by fitting the $10\,000$ noise realizations and the input spectrum (the ideal case of perfect match, i.e. ratio 1, is represented by the red dashed line). The red solid lines mark the $1\; \sigma$ relative errors of the input spectrum for an exposure time $t_{exp}=100\;sec$.}
    \label{fig_case1}
\end{figure*}

In the left panels of Fig.~\ref{fig_case1} we show the 2D histograms of the $10\,000$ fitted parameters for the test case with a spot $\Delta T=300\; K$ cooler than the photosphere and with $ff=0.03$ for the three reference stars. The input value is represented by a red cross, while the median value of the fitted spot parameters is marked in green. The contour level encompasses the most probable $68\%$ values of the fitted parameters. 

We remark that the 2D distribution of the fitted parameters is such that the spot parameters are anti-correlated, i.e. the contrast temperature $\Delta T$ decreases with increasing filling factor. This is consistent with the fact that the effect of large temperature contrasts tend to compensate small filling factors.\\

For each combination of the retrieved spot parameters, the stellar spectrum $F_\lambda^{interp}$ associated with the best-fit parameters is obtained by linearly interpolating the original 2D grid of stellar spectra  and then comparing it to the unperturbed input spectrum $F_\lambda^{input}$ (with $ff=0.03$ and $\Delta T=300\; K$). The corresponding ratios $F_\lambda^{interp}/F_\lambda^{input}$ are compared to the 
relative noise (see Eq.~\ref{noise}) associated with the input spectrum, evaluated for an exposure time $t_{exp}= 100\;sec$ (on the right panels of Fig.~\ref{fig_case1}). This procedure allows us to verify if the method confidently recovers the stellar spectrum within the noise. We find that the uncertainty due to the adopted procedure is within the $1\;\sigma$ relative noise of the unperturbed spectrum in AIRS channels; in the range $0.50-1.95\;\mu m$, the uncertainty of the retrieved spectra is above the noise level. These results can be explained considering that small variations of the spot parameters with respect to the input parameters produce a greater variation of the stellar flux in the optical and in the near infrared ($0.50-1.95\;\mu m$), where the spot-star contrast is greater.\\

We tested a number of cases by using different spot parameters, obtaining results consistent to what is discussed above. The results are summarized in Table~\ref{confronto_param}, where we show the intervals around the median values corresponding to the 16th and 84th quantiles of the $10\,000$ retrieved spot parameters. If we compare these values for the same star, we find smaller uncertainties on the fitted parameter for higher contrast spots, i.e. when the spot is better detectable. If instead we compare the case of the three reference stars each with the same input spot, the uncertainties are larger for the star with the highest photospheric temperature; in fact, when the three stars have the same visible magnitude and the same spot temperature contrast and size, we can characterize the spot less precisely in the hottest star, where the relative contrast $\Delta T=\frac{T_{*}-T_s}{T_*}$ is smaller. We verified that our fitting procedure is not strongly dependent on the starting conditions.\\

\begin{table}
    \centering
    \caption{Summary of the spots parameters obtained in specific cases, where the S/N ratio of the three reference stars is fixed ($V=9$, $t_{exp}=100\;sec$). $\Delta T_{input}$ and $ff_{input}$ are the input spot parameters. The last two columns show the median of the $10\,000$ retrieved spot parameters, the lower and upper values are the 16th and 84th quantiles.}
    {\def\arraystretch{2}\tabcolsep=5pt
    \begin{tabular}{crrrr}
    \hline
    Name star& $\Delta T_{input}(K)$ & $ff_{input}(\%)$ & $\Delta T(K)$& $ff(\%)$\\
    \hline

    \multirow{4}{*}{HD 17156} &  \multirow{2}{*}{300} & 3  & $250.0_{-109.7}^{+150.2}$ & $3.54_{-1.27}^{+2.61}$\\
    &                                                   & 10 & $282.1_{-46.6}^{+50.9}$   & $10.56_{-1.51}^{+1.95}$\\
                             & \multirow{2}{*}{1500}  & 3  & $1487.2_{-52.0}^{+50.7}$  & $3.00_{-0.05}^{+0.06}$\\
    &                                                   & 10 & $1496.0_{-14.5}^{+14.7}$  & $10.01_{-0.05}^{+0.05}$\\
    \cline{2-5}
    \multirow{4}{*}{HAT-P-11} &  \multirow{2}{*}{300} & 3  & $286.8_{-81.6}^{+90.1}$   & $3.09_{-0.62}^{+1.04}$\\
    &                                                   & 10 & $295.8_{-25.6}^{+23.9}$   & $10.11_{-0.63}^{+0.82}$\\
                             & \multirow{2}{*}{1500}  & 3  & $1504.7_{-19.0}^{+23.7}$  & $3.00_{-0.04}^{+0.04}$\\
    &                                                   & 10 & $1501.4_{-5.7}^{+7.0}$    & $10.00_{-0.03}^{+0.03}$\\
    \cline{2-5}
    \multirow{4}{*}{K2-21}  &   \multirow{2}{*}{300}  & 3  & $308.0_{-14.4}^{+13.9}$    & $2.93_{-0.09}^{+0.11}$\\
    &                                                   & 10 & $302.5_{-4.3}^{+3.9}$      & $9.93_{-0.09}^{+0.11}$\\
                            & \multirow{2}{*}{1500}   & 3  & $1512.1_{-25.8}^{+12.6}$   & $3.01_{-0.03}^{+0.03}$\\
    &                                                   & 10 & $1503.5_{-7.5}^{+3.8}$    & $10.02_{-0.03}^{+0.03}$\\
    \hline
    \end{tabular}
    }
    \label{confronto_param}
\end{table}

As noted in Section~\ref{Introduction}, because of the presence of spots, variations of the observed stellar flux may occur between observations taken at far apart times, due to the stellar rotation or changes in the stellar magnetic activity. Below we show the capability of the algorithm to retrieve the instantaneous spot parameters when comparing observations at different epochs, by focusing on the HAT-P-11 star whose stellar rotation causes an evident rotational modulation of the spot filling factor.Here we are not interested in simulating the details of the spots evolution, so we can test the method on a set of noisy spectra collected during a complete stellar rotation ($29.2$ days; \citealt{Bakos}) with a random normal distribution of filling factors ($<ff>=0.05$, $\sigma_{ff}=0.015$) and with a fixed spot temperature $T_s=4500.0K$ (\citealt{Morris}). The orbital period of the planet  is $4.89$ days, so during a rotation period we observe about 6 transits (\citealt{Rotation}). Subtracting the equivalent duration of six planetary transits (each transit lasts $\sim 2.1 hr$) from the  rotation period, we derive that during a complete stellar rotation it is possible to collect observations of the star unimpeded for about 28.7 days. We assume to observe each spectrum for an integration time of about an hour: this choice allows us to retrieve the spot parameters with a smaller uncertainty than assuming $t_{exp}=100\;sec$ but of course implies the assumption that  the  stellar flux does not significantly vary over the one hour exposition. We fix the visible magnitude of the star HAT-P-11 at  $V=9.6$.

\begin{figure*}
    \centering
    {\includegraphics[width=0.45\textwidth]{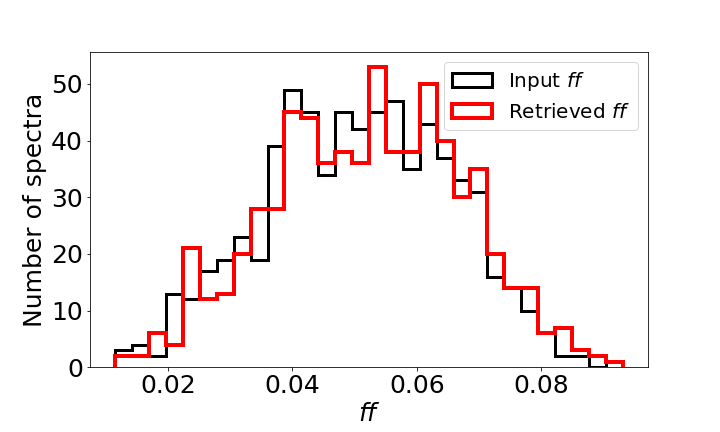}\quad
    \includegraphics[width=0.45\textwidth]{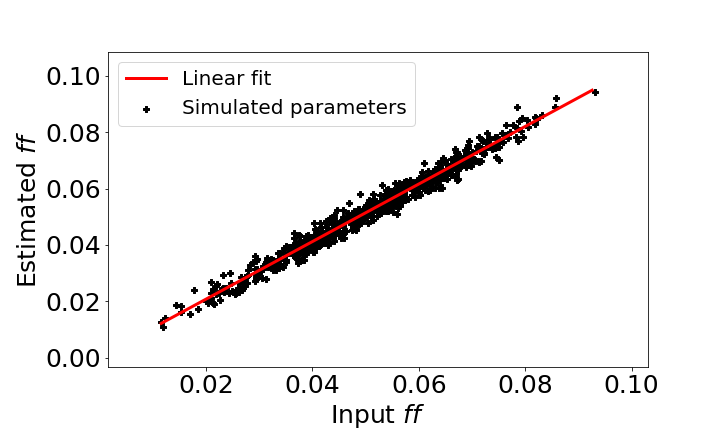}}
    \caption{\textit{Left panel}: histograms of the spots filling factors of the HAT-P-11 star. The black line is the input distribution, the red one shows the inferred distribution. Each input spectrum is integrated over one hour exposure. The star has a visible magnitude $V=9.6$. \textit{Right panel}: Relation between the input and the recovered filling factors. The red line is the linear fit (slope $m=1.02$, intercept $q=3.3\times10^{-4}$).}
    \label{Histograms}
\end{figure*}

Fig.~\ref{Histograms} shows that our fitting algorithm is able to reliably retrieve, from a statistical point of view, the distribution of the filling factors of the spots (left panel). The standard deviation of the retrieved spots temperatures is $\sigma_T=17.3\;K$. 
We remark that for each observation the filling factor of the spot is reliably recovered (see right panel of Fig.~\ref{Histograms}), so the out-of-transit observations may be used to sample the coverage of the spots on the observed hemisphere of the star. This analysis may allow to derive and to compare the filling factor of the spots between different observations, so to give information about the temporal evolution of the spots and about the stellar magnetic activity. 

\subsection{Robustness of the spotted star model}\label{model_limits}
\begin{figure*}
    \includegraphics[width=\textwidth]{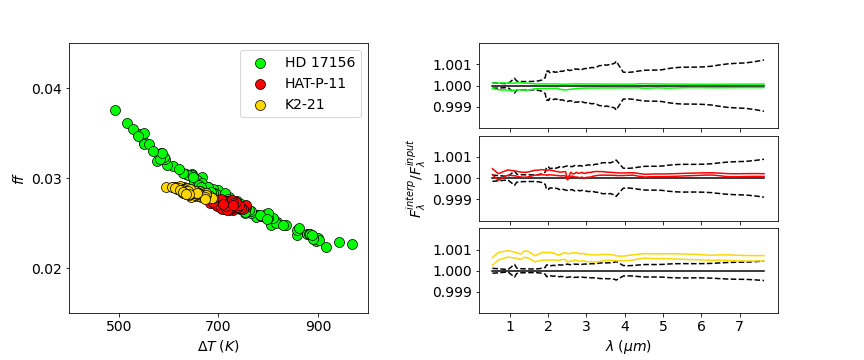}
    \caption{\textit{Left panel} - Scatter plot of the parameters got with a single-spot-model fitting and input stellar spectra generated instead assuming two spots with filling factors $ff_1=0.01$ and $ff_2=0.02$, and temperature contrasts $\Delta T_1=1000\;K$ and $\Delta T_2=500\;K$, respectively. Each color stands for a specific star (see legend). \textit{Right panels} - Ratio between the best-fit spectra obtained by fitting the 100 noise realizations and the input spectrum. The black solid line, i.e. ratio 1, represents the input spectrum. The black dashed lines mark the $1\;\sigma$ relative errors of the input spectra with an exposure time $t_{exp}=100\;sec$. The colored solid lines represent the 16th and 84th quantile of the 100 fitted spectra.}
    \label{fig_two_spot_model}
\end{figure*}

The model in Eq.~(\ref{model_out}) is based on the  assumption that the visible hemisphere of the star is dominated by the presence of a main spot with a given temperature, so it neglects that stellar spots may have different temperatures. In all the explored cases (summarized in Table~\ref{confronto_param}), the fitted distributions peak on the input spot parameters but this result may depend on the assumption that only one dominant spot has been taken into account. In order to understand if and how such an assumption affects our results, in the following we simulate a star with two spots having different sizes and temperatures and we fit the resulting simulated stellar spectrum with a single spot.  We set again the visible magnitude of the three stars at $V=9$  and we consider two spots with filling factors $ff_1=0.01$ and $ff_2=0.02$, and temperature contrasts $\Delta T_1=1000\;K$ and $\Delta T_2=500\;K$, respectively. For each star, we simulate 100 noisy spectra with $t_{exp}=100\;sec$ and we retrieve the pair of effective spot parameters. As shown in  Fig.~\ref{fig_two_spot_model} the fitted spot parameters (\textit{left panel}) correspond to intermediate values between those of two input spots, and the 16th and 84th quantiles of the 100 fitted spectra are comparable to the input noise level (except for the K2-21 star where the flux of the best-fit spectra are sistematically overestimated at all wavelengths).

\begin{figure*}
    \includegraphics[width=\textwidth]{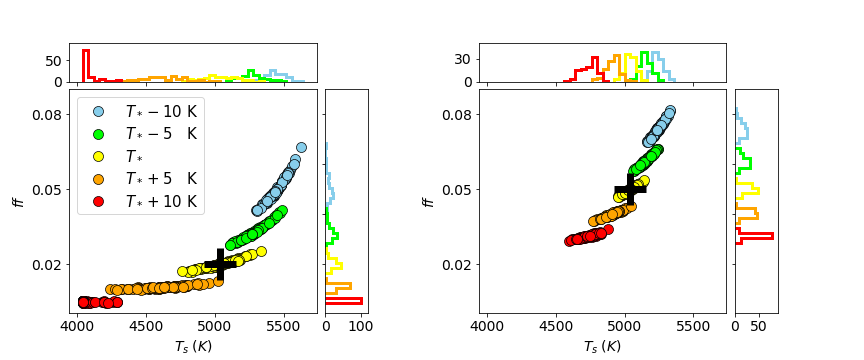}
    \caption{Scatter plots of the fitted spot parameters for stellar temperature values slightly different from the nominal value; the temperature difference  from the $6040\; K$ nominal value of HD 17156 is identified with a color for each set of simulations in the left panel legend. Each color pertains to $100$ noise simulations. The black cross marks the input temperature and filling factor of the spot. For each panel, the upper histogram shows the distribution vs. derived $T_s$, the one on the right the analogous distribution vs. derived $ff$. \textit{Left panel}: $ff = 0.02$, \textit{right panel}: $ff= 0.05$; for both panels the input temperature of the spot is $5040 \;K$.}
    \label{fig_error_on_T_eff}
\end{figure*}

Another limit of the model of Eq.~(\ref{model_out}) is that the stellar temperature $T_*$ is, in reality, not accurately known, so the model does not perfectly reproduce the true spectra. For instance, the photospheric temperature of the HD 17156 star is here set at $T_*=6040\;K$, but it is known with an uncertainty $\sigma_T\simeq24\;K$ (\citealt{Edwards}). We have explored how the uncertainty on the photospheric stellar temperature affects the determination of the spot temperature and of the filling factor. Fig.~\ref{fig_error_on_T_eff} shows the result of our algorithm when the temperature of the unperturbed photosphere is varied of a few degrees from the nominal value: for each change of $T_*$ from the nominal value, shown in the left panel legend, we generate a set of 100 noisy spectra corresponding to $T_s=5040\;K$, and $ff=0.02$ in the left panel and $ff=0.05$ in the right panel, again assuming an exposure time $t_{exp}=100\;sec$. The results show that both the spot temperatures and filling factors are overestimated if the photospheric temperature is lower than its nominal value, i.e. the best fit spectra correspond to a hotter and larger effective spot. On the contrary, if the temperature of the photosphere is higher than its nominal value, the spot parameters are both underestimated. 
Even if the effective temperature were precisely known, the model may still not perfectly fit, because our spectral models are necessarily incomplete (which is one of the reasons why the temperature values are uncertain), the surface gravity and metallicity of the star may also be uncertain and so on for the other parameters of the model.

\section{Retrieval of the planetary signal}\label{retrieval_planet}
In the following we simulate observations of planetary transits in the presence of spots in order to quantify and to correct the distortion of the planetary signal due to the spots in ARIEL observations. We propose a method to correct the transit depth variations for the cases of transits out of the the spots; we quantify the errors that we introduce by extending the same correction to transits over the spots. We consider the same systems that we have analyzed so far, this time taking also into account the occulting planets: HD 17156 b (a Jupiter), HAT-P-11 b (a Neptune) and K2-21 b (a super-Earth). The planetary parameters are shown in Table~\ref{planets}. We emphasize that in this Section we use the real visible magnitude of the three stars, shown in Table~\ref{input_param}. \\

In their study, \cite{Edwards} have analyzed a wide variety of target planets that ARIEL will observe during its 4-years mission in order to derive the science time required to detect a primordial atmosphere (i.e. H$_{2}$-He dominated with mean molecular weight $\mu=2.3 \;u$) on these targets for both transit and eclipse observations in the three ARIEL Tiers (with S/N$\geq$7). ARIEL's observations are based on a tiered approach and each Tier will answer different mission science questions and requirements (see \citealt{Tinetti_2018} for a more detailed description of ARIEL's tiering approach), therefore the three spectral channels will have a different resolution on each Tier (summarized in Table \ref{tier}). 

In addition to the spectroscopic measurements, ARIEL will also provide the stellar flux measurements in the photometric channels for each Tier.
\begin{table}
    \centering
    \caption{Resolution of each spectral channel in Tiers devoted to transit observations. The spectral range for each channel is shown in Tab. \ref{Instr}.}
    \begin{tabular}[h!]{lccc}
    \hline
    Instrument Name & Tier 1 & Tier 2 & Tier 3 \\
                    & (R)    & (R)    & (R) \\
    \hline
    VISPhot  &   1      &   1       &   1     \\
    FGSPrime &   1      &   1       &   1     \\
    FGSRed   &   1      &   1       &   1     \\
    NIRSpec  &  $\sim$1 &  $\sim$10 &  $\sim$20\\
    AIRS Ch0 &  $\sim$3 &  $\sim$50 &  $\sim$100\\
    AIRS Ch1 &  $\sim$1 &  $\sim$10 &  $\sim$30\\
    \hline
    \end{tabular}
    \label{tier}
\end{table}
Here, we focus on transits observations at Tier 2 resolution, as Tier 2 spectroscopic measurements will be crucial for detecting molecular absorption features in the transmission spectrum of the planetary atmosphere. The number of transits simulated for each transiting system are shown in Table~\ref{planets}.\\

\begin{table}
    \centering
    \small
    \setlength\tabcolsep{4pt}
    \caption{Radius, mass, equilibrium temperature, transit length $\tau$ and impact parameter $b$ of the three analyzed planets, from \citet{Edwards} catalogue. $n$ is the number of transits needed to observe a primordial atmosphere at Tier 2 resolution with S/N=7.}
    \begin{tabular}[h!]{ccccccc}
    \hline
    Planet     & Radius  & Mass & $T_{eq}$ & $\tau$ & $b$ & $n$\\
    &($R_{\oplus}$)&($M_{\oplus}$)& ($K$) &  ($hr$)& &\\
    \hline
    HD 17156 b &  12.1 & 98.3 & 904.0  &  4.2  &  0.89  & 2\\
    HAT-P-11 b &  4.3  & 17.1 & 847.8  &  2.1  &  0.5   & 3\\
    K2-21 b    &  1.8  & 3.9  & 647.2  &  2.6  &  0.45 & 184\\
    \hline
    \end{tabular}
    \label{planets}
\end{table}

\begin{figure}
    \centering
    \includegraphics[width=0.45\textwidth]{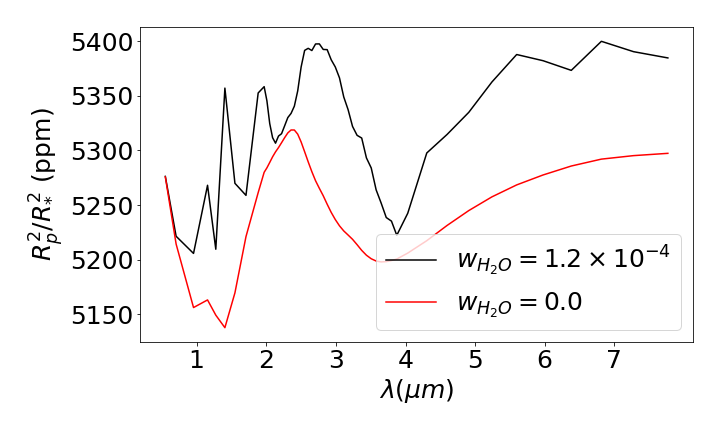}
    \caption{Examples of transmission spectra of the planet HD 17156 b at ARIEL Tier 2 resolution. The abundance ratio between He and H$_2$ is 0.17, while the water volume mixing ratio $w$ is specified in the legend.}
    \label{Tr_spectrumG_spat}
\end{figure}

For each planet, we generate the synthetic transmission spectrum $\varepsilon_\lambda$ with the TauRex code (\citealt{TauREx}), by assuming that the three selected targets have primordial atmospheres, with an abundance ratio between H$_{2}$ and He typical of a primordial atmosphere. We assume an isothermal temperature profile (the equilibrium temperature is shown in Table~\ref{planets}), that the planetary atmosphere extends for five scale heights (\citealt{sing_2018, description_observation}), and we set the pressure at the bottom of the atmosphere to $1\; bar$ for both terrestrial and gaseous planets\footnote{Generally, for gaseous planets the pressure at the surface of the planet is $p_{0}\sim 1-10\; bar$, depending on the transparency of the atmosphere, \citealt{description_observation}.}. Furthermore, for each target we generate another transmission spectrum adding traces of water (with volume mixing ratio $w\sim 10^{-4}$). 
The presence of such a small amount of water does not significantly affect the mean molecular mass of the particles of the atmosphere, therefore we can assume that the transit observation times of the three targets of planets are not sensibly different from those estimated by \cite{Edwards} for primordial atmospheres. However the water molecules are by far the most important (among the three considered) in the planetary transmission spectrum since their presence causes the appearance of molecular bands. Fig.~\ref{Tr_spectrumG_spat} shows the synthetic transmission spectra generated for the HD 17156 b planet: the black solid line shows the contribution of H$_{2}$ and He to the transmission spectrum, while the red solid line shows the resulting spectrum in the presence of water.\\

We stress that in this work we are not interested in modeling the details of the planetary transmission spectrum or of the chemical atmospheric composition, but we  want to test if our method is able to correctly derive the planetary transmission spectrum in the presence of spots, and whether our method can identify the presence of small traces of water in the planetary atmosphere.

\subsection{Transits out of the spots}
We simulate planetary transits out of the spots, exploring two possible configurations of the spot parameters: $\Delta T=300\;K$, $ff=0.03$ and $\Delta T=1000\;K$, $ff=0.10$. For each planetary system, we generate $n$ observations ($n$ is reported in Table~\ref{planets}), by simulating both cases with primordial atmospheres, and with traces of water. A single observation consists of an out-of-transit spectrum $F_\lambda^{out}$ and an in-transit spectrum $F_\lambda^{in}$, both integrated over an exposure time equivalent to the transit length $\tau$ (see Table~\ref{planets}). We can integrate the transit out of the spot over the transit duration since we are assuming that the planet always blocks the same stellar flux regardless of its position on the stellar disk, i.e. we are ignoring the limb darkening effects. Since ARIEL noise is dominated by the photon noise, using such a long integration time reduces the relative noise of the observed spectra and allows us to recover the spot parameters from out-of-transit observations with a small uncertainty. Each out-of-transit stellar spectrum is built as explained in the Section~\ref{sim_out_transit}, while the in-transit spectrum is built as in Eq.~(\ref{model_in_transit}), where $\varepsilon_\lambda$ is the transmission spectrum of the considered target of planet, $T_s$ and $ff$ are the same spot parameters used to generate the out-of-transit spectrum and $g_\lambda =0$, as we are simulating transits out of the spots.\\

In this work, the main assumption is that the pattern of the spot does not change between the out-of-transit and the in-transit observation. However, if we consider observations taken at different times, the spot coverage on the stellar hemisphere exposed to the observer may vary due to the stellar rotation or changes in the stellar magnetic activity. In order to take into account the possible changes of the filling factor of the spots between the observations, we randomly generate the spot filling factor in each observation from Gaussian distributions of filling factors peaked at the nominal values ($<ff>=0.03$ in the first case, and $<ff>=0.10$ in the second case) and with standard deviation $\sigma_{ff}=0.02$.\\

\begin{figure*}
    \centering
    \includegraphics[width=\textwidth]{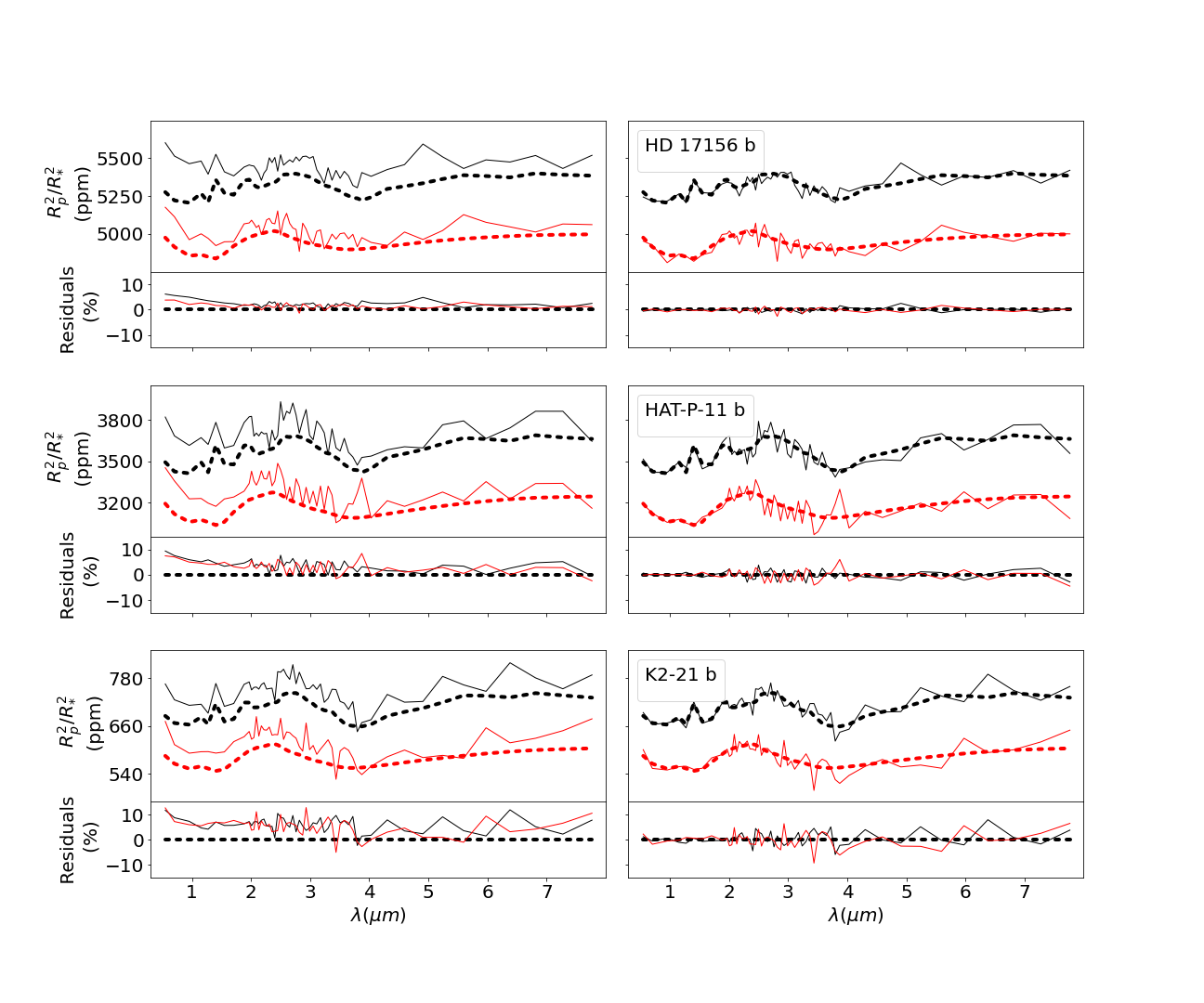}
    \caption{Retrieved planetary spectra, uncorrected (\textit{left panel}) and corrected (\textit{right panel}) for the presence of the spot, averaged over the $n$ observations ($n$ is reported in Table~\ref{planets}). In each simulated observation, the spot temperature contrast is $\Delta T=1000\;K$, while the filling factor is randomly extracted from normal distributions peaked at $<ff>=0.1$ and having $\sigma_{ff}=0.02$. The dashed line are the input planetary transmission spectra $\varepsilon_\lambda$, while the continuous lines are the retrieved spectra, without water (black lines) and with $10^{-4}$ mixing ratio of water (red lines). The transmission spectra with water are shifted down for clarity. For each case the residuals are expressed as percentage deviations from the input transmission spectrum. The spectra are degraded at the Tier 2 resolutions.}
    \label{fig_case2_with_planet}
\end{figure*}

If we estimated the planetary spectrum as $\delta_\lambda=\frac{F_\lambda^{out}-F_\lambda^{in}}{F_\lambda^{out}}$ in each observation  (i.e. without correcting for the presence of the  non occulted spot) averaging on the number $n$ of the observations, we would overestimate the extracted planetary signal, and the effect would depend on the specific combination of the spot parameters (see left panels of Fig. \ref{fig_case2_with_planet}). However, as discussed in Section~\ref{primary_transit}, if the planet transits out of the spots and the spots pattern does not significantly vary in the single observation, we may extract the planetary signal $\varepsilon_\lambda$ in each observation as follows:
\begin{equation}
    \label{extraction_planet}
    \varepsilon_\lambda=\frac{F_\lambda^{out}-F_\lambda^{in}}{F_\lambda(T_*)}
\end{equation}
where $F_\lambda(T_*)$ is the spectrum of the non-spotted photosphere that we retrieve from the corresponding out-of transit observation; in particular, $F_\lambda(T_*)$ is derived from Eq.~(\ref{model_out}), where $ff$ and $T_s$ are the spot parameters estimated from $F_\lambda^{out}$ as described in Section~\ref{sim_out_transit}, and $F_{\lambda}(T_{s})$ is interpolated from our grids of spots spectra. Therefore our method can derive both the characteristics of the spots and the spectrum of the non-spotted photosphere, thus removing the effects of the spots. Each observation is simulated at Tier 3 resolution, and then $F_\lambda^{out}$, $F_\lambda^{in}$ and $F_\lambda(T_*)$ are degraded at Tier 2 resolution.

The panels of Fig.~\ref{fig_case2_with_planet} show the retrieved planetary signals, uncorrected (left panel) and corrected (right panel) for the presence of the spot, averaged over the $n$ observations, with $\Delta T=1000\;K$. In these cases, neglecting the presence of non-occulted spots leads to significant overestimates of the inferred planetary spectrum (see left panels of Fig.~\ref{fig_case2_with_planet}), while our  method removes the systematic overestimate of the planetary signal, thus recovering the planetary spectrum within the noise. Furthermore, the method can distinguish the case of primordial atmospheres from the one with traces of water. On the other hand, we have verified that the correction becomes marginal at a lower spot contrast of $\Delta T=300\;K$ and therefore not necessary for these cases. The standard deviations $\sigma$ of the residuals of the recovered planetary signals (summarized in Table~\ref{variance}) show that the method allows to recover the planetary signal within the $3.3\%$.

\begin{table}\footnotesize{}
    \centering
    \caption{Standard deviations of residuals of the recovered planetary signals for $ff$ and $\Delta T$ input, corrected for the presence of non-occulted spots, in the explored cases. $w$ is the water mixing ratio.}
    \scalebox{0.9}{\begin{tabular}{crrrrr}
    \hline
    $w$       &  $ff$     & $\Delta T(K)$ & HD 17156 b &  HAT-P-11 b & K2-21 b\\
    \hline
    \multirow{2}{*}{0}  &   0.03     & $300 $     & $0.8\%$    & $1.8\%$   & $2.9\%$\\ 
                        &   0.10     & $1000 $    & $0.7\%$    & $1.6\%$   & $2.6\%$\\
    \cline{2-6}
    \multirow{2}{*}{$1.2\times 10^{-4}$}  &   0.03     & $300 $     & $0.8\%$    & $2.1\%$   & $3.3\%$\\ 
                                          &   0.10     & $1000 $    & $0.8\%$    & $2.0\%$   & $2.9\%$\\
    \hline
    \end{tabular}}
    \label{variance}
\end{table}

\begin{figure}
    \centering
    \includegraphics[width=\columnwidth]{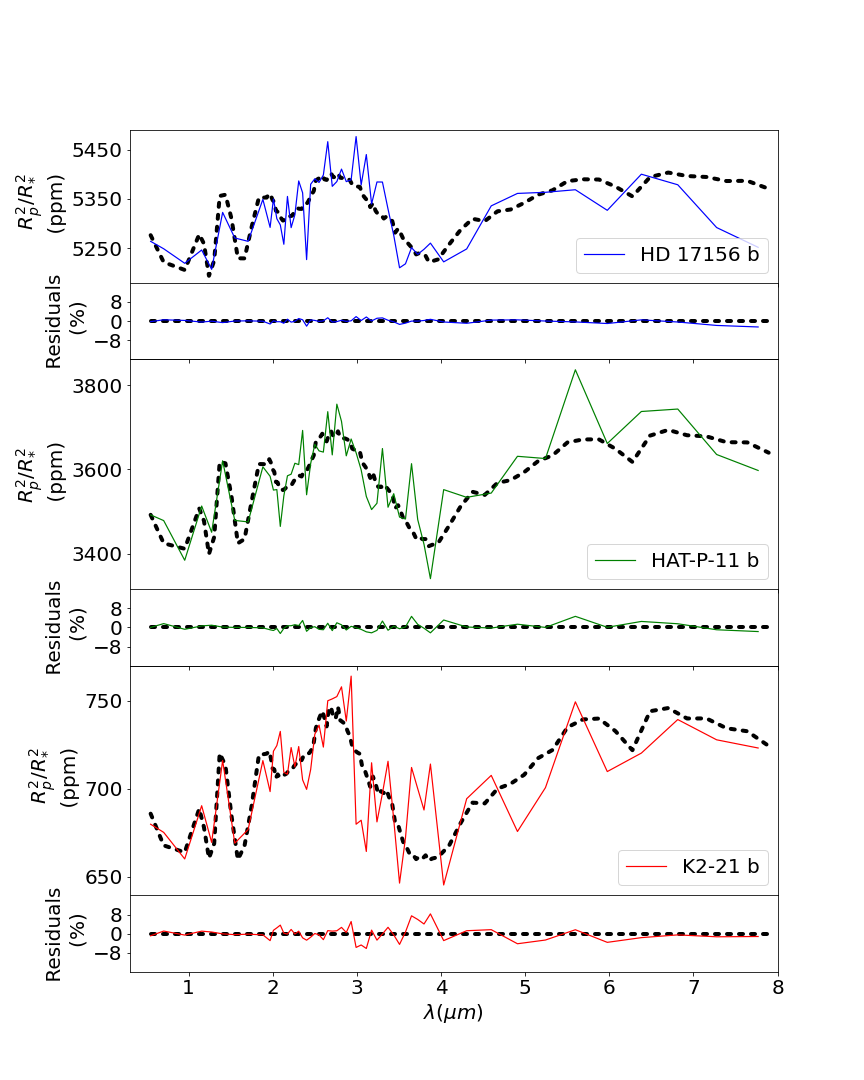}
    \caption{Examples of transmission spectra (simulated with traces of water) corrected for the presence of a spot when the star has two spots, with temperature contrasts $\Delta T_1=1000\;K$ and $\Delta T_2=500\;K$, respectively, and filling factors randomly extracted from Gaussian distributions with mean $<ff>=0.02$ and standard deviation $\sigma_{ff}=0.02$ in each observation. Each planetary spectrum is averaged over $n$ observations (Table~\ref{planets}) and degraded at Tier 2 resolution. For each case, the percentage residuals are shown.}
    \label{planetG_wiht_two_spot}
\end{figure}

As discussed in Section~\ref{model_limits}, we are using a simplified model, assuming a nominal stellar temperature and a single dominant spot. In the previous section we have shown that if the true spot distribution is more complex and the stellar temperature is known with an uncertainty of few tens of degrees, we rather obtain a set of effective spot parameters that describe the spectrum reasonably well.
Below we show that our method provides reasonable transmission spectra even if the star  has two different spots. As in the previous case we assume that in the single observation the spots configuration does not change but can be different during different transits. We simulate $n$ observations for each target planet; also for each out-of-transit observation we derive the effective spot parameters and the spectrum of the unperturbed photosphere in order to retrieve the planetary transmission spectrum. The inferred transmission spectrum is averaged over the $n$ observations. In Fig.~\ref{planetG_wiht_two_spot} we show the specific case where we fit a one-spot model to a star with two spots having temperature contrasts $\Delta T_1=1000\;K$ and $\Delta T_2=500\;K$, respectively, and filling factors randomly extracted from Gaussian distributions with mean  $<ff>=0.02$ and standard deviation $\sigma_{ff}=0.02$ in each observation: the planetary transmission spectrum is recovered with a spread of a few percent from the input value, analogously to the case in which the star has only one spot.
\\

\begin{figure}
    \centering
    \includegraphics[width=\columnwidth]{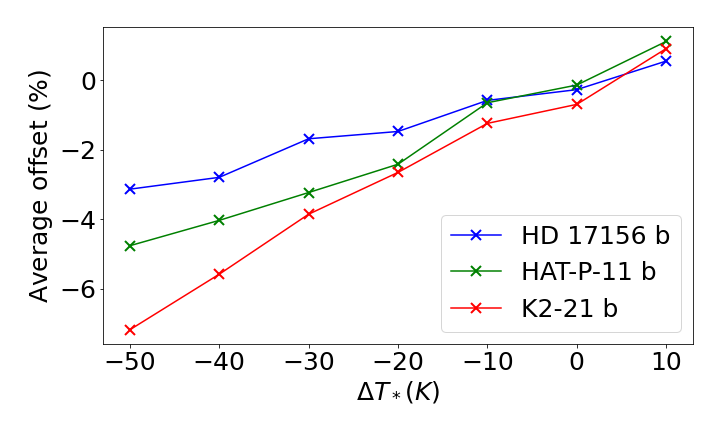}
    \caption{Average percentage offsets of the retrieved planetary spectra in ARIEL photometric channels for different input photospheric temperatures. $\Delta T_*$ is the error of the photospheric temperature. The star is simulated as having a spot with temperature $1000\;K$ lower than the nominal photospheric temperature and with filling factor randomly extracted from a Gaussian distribution with mean $<ff>=0.02$ and $\sigma_{ff}=0.02$. Each point in the plot derives from the average of the planetary signal over $n$ noisy simulations (Table~\ref{planets}).}
    \label{planetG_wiht_error_on_t_eff}
\end{figure}

With a similar approach, we determine the error we introduce on the planetary spectrum extraction when the photospheric temperature $T_*$  is known with an uncertainty of few tens of degrees. We focus on the case with a spot having a temperature contrast of $1000\;K$; for each transiting system we simulate $n$ observations, each with a filling factor randomly extracted from a Gaussian distribution with mean $<ff>=0.02$ and $\sigma_{ff}=0.02$. We extract the planetary transmission spectrum from each transit, and  average over the $n$ observations. We find that the size of the errors of the extracted planetary signal are correlated with the error of $T_*$, i.e. the planetary transmission spectrum is underestimated when the photospheric temperature is underestimated and vice versa, with more evident effects in the bluest photometric channels. As an example, in Fig.~\ref{planetG_wiht_error_on_t_eff} we report the average offset of the retrieved planetary spectra (simulated with traces of water) in the photometric channels for different errors $\Delta T_*$ of photospheric temperatures. The results show that the planetary transmission spectrum is  recovered within $1.0\%$ for photospheric temperatures errors within $10\;K$ (comparable with the spread of the planetary spectrum recovered for $\Delta T_*=0$), while the signal is significantly distorted for greater errors of $T_*$. Therefore, our methods provides the best results if the stellar parameters are accurately known.

\subsection{Transits over the spots}
When the planet crosses the spot during the transit, a bump will occur in the transit light curve. If the bump is clearly visible, we are able to separate it from the remaining transit, and we can just correct the points of the transit out of the spot crossing with the method used in the previous paragraph. However, in many cases, due to the low S/N ratio, the bump is not evident in the light curve. In this work, we are not interested in developing a method to correct transits over the spots, but we will evaluate the impact on the planetary signal extraction if we analyze the case of a transit over the spot with the same method used to correct non-occulted spots.\\

For simplicity we assume that the spot projection on the stellar disk is a circle\footnote{Generally, the projection of a circular spot on the stellar disk is an ellipse whose semi-major axis is the radius of the circular spot and semi-minor axis depends on the angle between the line of sight and the normal at the center of the spot. A real spot, however, can have a more complicated shape.} with radius $R_{spot}=\sqrt{ff}R_*$ and it is in the transit chord, so both the spot and the planet have the same impact parameter. We bin each transit observation with $t_{exp}=100\;sec$ exposures, so if $\tau$ is the transit length, we will have $N=\tau/t_{exp}$ points per transit. The fraction $k$ of points per transit contaminated by the spot crossing is $k=\frac{2\left(R_p+R_{spot}\right)}{L}$, where $L$ is the transit chord length which is a function of the impact parameter $b$ (reported in Table~\ref{planets}). In the hypothesis of circular orbits and that the planet is completely inside the stellar disk during the transit, $L$ can be expressed as (\citealt{Seager}):
\begin{equation}
    L=2\sqrt{(R_*-R_p)^2-(bR_*)^2 }
\end{equation}
For the star HD 17156, for instance we will have $N=149$ data points per transit and if we consider a spot with filling factor $ff=0.001$ in the transit chord, the crossing over the spot will affect a fraction $k\sim 2/5$ of the transit data points. This fraction will be larger for larger spot sizes.\\

During the transit, the spectrum $F_\lambda^{in}$ of each exposure has to include the flux coming from the spot occulted by the transiting planet, as described in Eq.~(\ref{model_in_transit}), that takes into account the wavelength dependence of the planet radius. In the case of a circular spot we obtain:

\begin{equation}
\begin{split}
    \label{intersection}
    g_\lambda \varepsilon_\lambda & = \frac{1}{\pi R_*^2}\left[ R_\lambda^2\arccos{\left(\frac{d_{1\lambda}}{R_\lambda}\right)}-d_{1\lambda}\sqrt{R_\lambda^2-d_{1\lambda}^2}\right. \\ 
    & \left.+R_{spot}^2\arccos{\left(\frac{d_{2\lambda}}{R_{spot}}\right)}-d_{2\lambda}\sqrt{R_{spot}^2-d_{2\lambda}^2}\right]\\
\end{split}
\end{equation}
where $R_\lambda$ is the planetary radius as a function of wavelength, while $d_{1\lambda}$ and $d_{2\lambda}$ are respectively:

\begin{equation}
\begin{split}
    d_{1\lambda} & =\frac{R_\lambda^2+d^2-R_{spot}^2}{2d}\\
    d_{2\lambda} & =d-d_{1\lambda}\\
\end{split}
\end{equation}
$d$ is the distance between the spot and planet centers. The detailed derivation of Eq.~(\ref{intersection}) is reported in Appendix~\ref{derivation}. \\

Generally, in the presence of spot crossing events, we have to distinguish between two possible configurations, according to the relative size of the spot and planet:
\begin{itemize}
    \item $ff\leq\left(\frac{R_p}{R_*}\right)^2$: the spot is smaller than the planet, so a complete occultation of the spot during the transit will occurs when $g_\lambda \cdot \varepsilon_\lambda=ff$. In general, the fraction $g_\lambda$ will span the interval $\left[ 0,\frac{ff}{\varepsilon_\lambda}\right]$ during the transit;
    \item $ff>\left(\frac{R_p}{R_*}\right)^2$: the spot is larger than the planet, so $g_\lambda$ will span the interval $\left[ 0,1\right]$, where $g_\lambda=1$ occurs when the planet is entirely over the spot.
\end{itemize}

\begin{figure*}
    \centering
    {\includegraphics[width=0.48\textwidth]{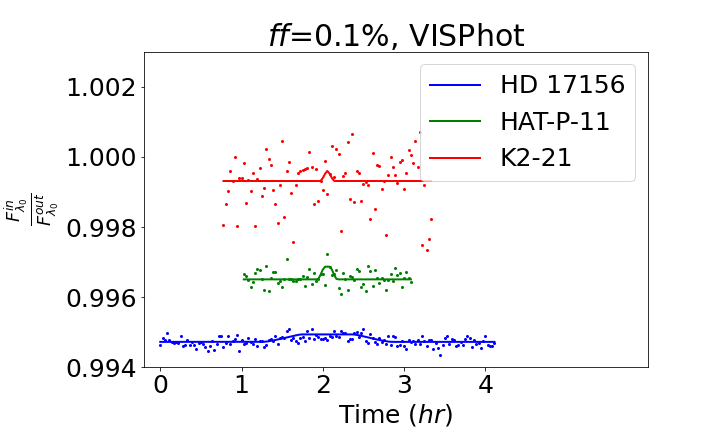}\quad
    \includegraphics[width=0.48\textwidth]{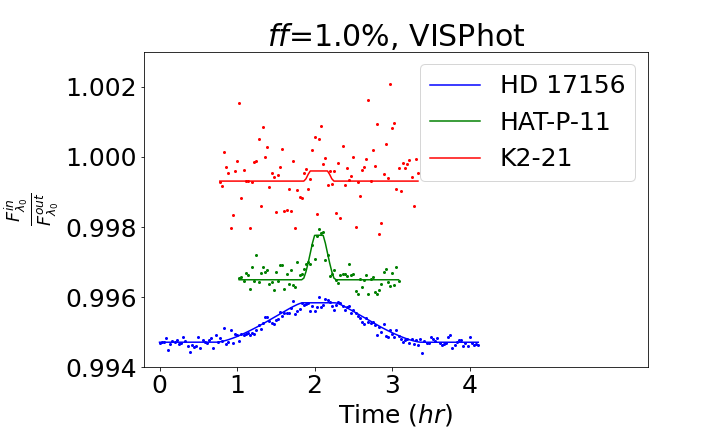}}
    \caption{Transits light curves (omitting the out of transit, and ingress/egress of the planet) of the three analyzed planetary systems in the first ARIEL photometric channel, where the crossed spot is at half the transit time with the same impact parameter of the transiting planet. The input transmission spectrum of each target contains traces of water. The continuous lines mark the cases without noise. The filling factors of the spot is shown in the header of the plot, and its temperature is $300\;K$ cooler than the photosphere.}
    \label{light_curve}
\end{figure*}

Here we simulate transit observations over the spots, by exploring both the possible configurations. In particular, for each of the three planets studied above we simulate transit observations where the crossed spot has two possible temperature contrasts ($\Delta T=300\;K$ and $\Delta T=1000\;K$), and different filling factors in the interval $[0.001, 0.02]$\footnote{Due to the high impact parameter of both the planet and the spot, for the HD 17156 b system, a circular spot can be entirely in the visible stellar hemisphere only if its filling factor is smaller than $0.012$. Furthermore, for the K2-21 b, the planet-to-star areas ratio is $6\times 10^{-4}$, so transits over spots smaller than the planet can be analyzed by extending the explored spot filling factor interval towards smaller values.}. For each couple of the spot parameters we simulate a single observation where the in-transit observation is binned with $100\;sec$ exposures and the out-of-transit observation is integrated over a time equivalent to the transit length $\tau$. We simulate primordial atmospheres with traces of water. As in the previous Section, we work in the hypothesis that the out-of-transit and the in-transit filling factors do not differ, since we are focusing on the same observation. 
As an example, Fig.~\ref{light_curve} shows two light curves for each of the three transiting systems for two specific configurations of the spot parameters. We can see that for a filling factor smaller than 0.001 (and spot temperature contrasts $\Delta T$ lower than $300\;K$) the occultation of the spot is not visible in the three light curves; for $ff=0.01$ the bump due to the spot crossing is instead clearly visible for the HD 17156 and HAT-P-11 stars, while it  is, also in this case, comparable to the noise level for the K2-21 star. This is due to the low S/N ratio of the observation of this star (its visible magnitude is $V=12.8$).\\

If we choose not to correct for the presence of the spot (or rather we cannot because we have not realized its presence), the transit depth will be overestimated in the data points where the spot is not occulted and underestimated in the points where the spot is crossed by the transiting planet, with an average effect depending on the specific case. If instead we adopt the method described in the previous Section to correct both occulted and non-occulted spots, we will obtain:
\begin{equation}
    \label{correction_over_spots}
    \frac{\Delta F_\lambda}{F_\lambda(T_*)}=\frac{F_\lambda^{out}-F_\lambda^{in}}{F_\lambda(T_*)}=\varepsilon_\lambda\left(1-g_\lambda\frac{\delta F_\lambda}{F_\lambda (T_*)}\right)
\end{equation}
where $\delta F_\lambda=F_\lambda(T_*)-F_\lambda(T_s)$. The term between parentheses in Eq.~(\ref{correction_over_spots}) is difficult to determine,  because $g_\lambda$ is hard to estimate. However, neglecting this term would  lead us to underestimate the planetary transmission spectrum extracted 
as $\frac{\Delta F_\lambda}{F_\lambda(T_*)}$ by an amount depending on the spot temperature, $g_\lambda$ and $F_\lambda(T_*)$. In the following we will determine this underestimate for various couples of filling factor and temperature values, in order to understand its impact with respect to the noise level and whether it introduces a significant chromatic dependence. 
\\

\begin{figure*}
    \centering
    {\includegraphics[width=0.48\textwidth]{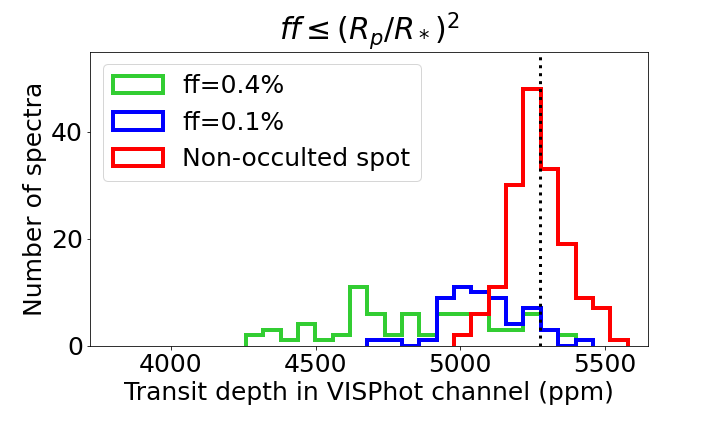}\quad
    \includegraphics[width=0.48\textwidth]{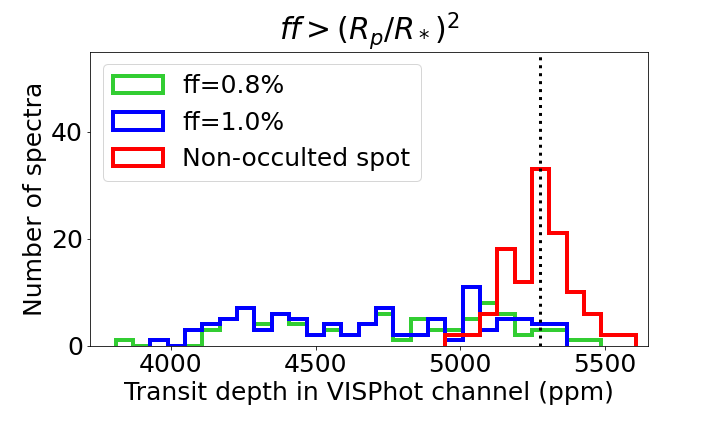}}
    \caption{Histograms of the transit depths retrieved from Eq.~(\ref{correction_over_spots}) in all the exposures for the first ARIEL photometer. The illustrated case refers to the planetary system HD 17156 b, with a stellar spot $300\;K$ cooler than the photosphere. Red color identifies the transit data points if the spot is not crossed in any of the two simulated transits, while the other colors identify the transit data points when the spot is crossed, for different star-disk filling factors. The vertical dotted black line is the input transit depth. On the left the case with $ff\leq\left(\frac{R_p}{R_*}\right)^2$, on the right the case with $ff>\left(\frac{R_p}{R_*}\right)^2$.}
    \label{Hist_flux_planet}
\end{figure*}

As an example, in Fig.~\ref{Hist_flux_planet} we show the distributions of the retrieved transit depth in all the exposures, derived from Eq.~(\ref{correction_over_spots}), in ARIEL VISPhot channel, for the planetary system HD 17156 b, by simulating two transits with different spot filling factors and spot temperature contrast $\Delta T=300\;K$. Out of the spot crossing (\textit{red distribution}) we are able to correct the transit depth since the distribution of the retrieved results is narrow with a well defined peak at the input value (\textit{black dotted line}), while during the spot crossing the signal is systematically underestimated and the retrieved results are distributed far from the true value (\textit{blue and green distributions}). Moreover, if the spot filling factor is smaller than the planet-star areas ratio (left panel of Fig.~\ref{Hist_flux_planet}) and is different between the two transits, we clearly distinguish the distributions of the retrieved transit depths during the spot crossings in the two cases (\textit{blue and green distributions}). Therefore, the underestimates at each exposure are measurable, at least for this planet and with the chosen combinations of spot parameters, and if the distribution of the recovered transit depths is asymmetrical (with a tail towards smaller values) we can identify and remove the transit data points affected by the spot crossing.\\

\begin{figure}
    \centering
    \includegraphics[width=\columnwidth]{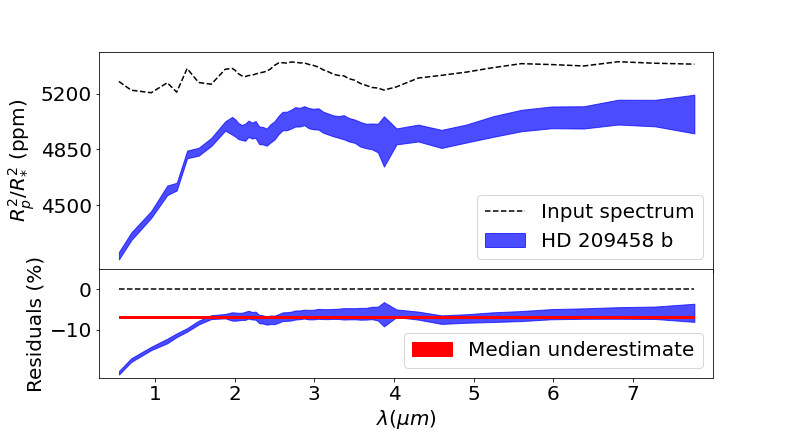}
    \caption{\textit{Upper panel:} The blue region marks the values within one sigma of the transmission spectra of HD 17156 b derived from \numprint{1000} noisy transit observations, where the crossed spot has $ff=0.01$ and $\Delta T=1000\;K$. The black dashed line is the input transmission spectrum. \textit{Bottom panel:} the blue region shows the residuals within-one sigma, while the narrow red region indicates the median underestimates within-one sigma computed over the entire band.}
    \label{example}
\end{figure}

For each couple $ff$, $\Delta T$, we simulate \numprint{1000} transit observations and we extract the planetary spectrum from each transit as described above. A comparison between the extracted planetary spectra and the input planetary signal shows that the retrieved planetary spectrum is systematically biased towards smaller values. We quantify this systematic underestimate in each of the \numprint{1000} simulations by evaluating the median value $M$ of the residuals of the retrieved planetary spectrum computed over the entire band. As an example, Fig.~\ref{example} shows the retrieved planetary simulated spectra of HD 17156 b for a crossed spot having $ff =0.01$ and $\Delta T=1000\;K$: the blue region in the upper plot delimits the values within one sigma of the \numprint{1000} derived spectra, while the blue region in the bottom panel shows the corresponding residuals, while the narrow red area indicates the one-sigma median ($M$) residuals. 
The residuals exhibit a systematic underestimate mainly below $2\mu m$, showing a chromatic effect particularly relevant at shorter wavelengths, where the spot-photosphere contrast is larger. The systematic offset is relevant for the brightest targets (i.e., HD 17156 and HAT-P-11), for which the effect is larger than the statistical noise, while it is less evident for the lower S/N spectrum of the super-Earth K2-21 b where the statistical fluctuations are larger than the offset.

\begin{figure}
    \centering
    \includegraphics[width=\columnwidth]{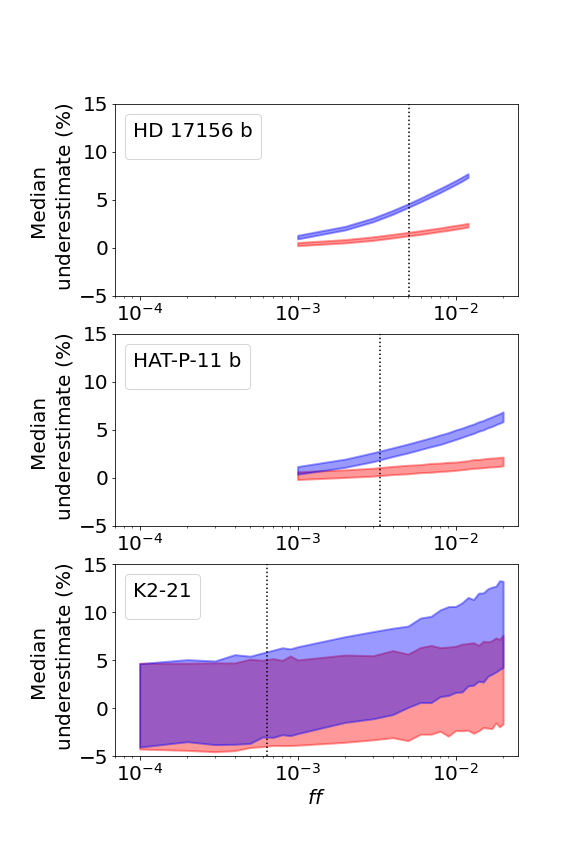}
    \caption{The colored regions show the median underestimates within-one sigma obtained from planetary spectra extracted from \numprint{1000} noisy transit observations, for different sizes and temperatures of crossed spot. The red color corresponds to $\Delta T=300\;K$, the blue to $\Delta T=1000\;K$. The black dashed vertical lines represent the planet-to-star area ratios.}
    \label{transits_over_spots}
\end{figure}

The extension of this analysis to the three reference planetary systems allow us to identify the median residuals for specific couples of $ff$ and $\Delta T$. The results are shown in Fig.~\ref{transits_over_spots}, where each colored region marks the values within one sigma of the median residuals derived from the \numprint{1000} simulations for various combinations of the spot parameters.
This analysis shows that even in cases of bright stars, for small values of spot filling factor $ff$ and low temperature contrasts $\Delta T$, the systematic bias of the retrieved planetary spectrum may be neglected since the median bias is of about $1\%$, i.e. comparable to the spread of the retrieved signal for non-crossed spots. Conversely, for high contrast spots, the underestimates introduced by the method are larger but in these cases the bumps are visible in the light curve and they can be removed from data, thus limiting the planetary spectrum extraction to the transit data points where the spot is not occulted. The effects are never measurable for K2-21 b.

\section{Summary and discussion}\label{Summary} 
We have developed a method to extract the atmospheric transmission spectrum for systems with transiting planets in the presence of non-occulted stellar spots by modeling the spot temperature and size from out-of-transit observations. The method is proposed for the ARIEL mission and it is tested with three ARIEL's potential targets of planetary systems, taken from the \cite{Edwards} catalogue: HD 17156 b, HAT-P-11 b, K2-21 b. For each target, we have generated a set of out-of-transit spectra for stars with one-dominant spot by linearly combining the spectrum of the unperturbed photosphere with spectra corresponding to pieces of photosphere at cooler temperatures, taken from BT-Settl models (\citealt{Baraffe}), and different filling factors (see Eq.~\ref{model_out}), thus mimicking the presence of one spot (for different physical and geometric parameters) on the stellar surface and taking into account the instrumental properties.\\

We have developed an algorithm to fit the noisy spotted star spectra and to recover the spot temperature and filling factor; in order to test the method we have fixed the visible magnitude of the three stars at $V=9$  for uniformity of S/N ratio. The results of our algorithm show that we reliably recover the spot's parameters. In particular, in the cases of low contrast between the photosphere and the spot, the studied cases show that the effect of the noise is larger than the effect of the spot, therefore the uncertainties of the fitted parameters are large; vice versa, for a high spot-star contrast the fitted spot parameters are less uncertain and span smaller intervals of $ff$ and $T_s$. The bluest photometric channels contribute most to the retrieval of the spot parameters, because the spot contribution to the stellar spectrum is larger at shorter wavelengths. Furthermore, for low temperature contrasts and small filling factors (i.e. $\Delta T<300.0\;K$ and $ff<0.03$), the fitted spot parameters show a significant dependence on the initial condition of the fitting algorithm for the star HD 17156, making very hard to detect the presence of small spots with low temperature contrast in this type of star. However we have verified that in these cases the spot-related influence on the planetary transmission spectrum extraction may be ignored, because comparable to the noise level.\\

With this method, the analysis of the out-of-transit spectra of each ARIEL's target will lead to the systematic characterization of photospheric activity due to the stellar spots in stars of different spectral types, providing valuable information about the temperature and size of spots on the stellar surface. In addition, the method will help to characterize the temporal evolution of the spots on different kinds of stars when observations at different epochs are compared, giving important information on some characteristics of magnetic activity as a function of star properties. \\

For each target, we have analyzed the effect of the spots on transit simulations and we have evaluated in which regimes the presence of non-occulted spots significantly affect the extraction of the planetary signal. To simulate the transits observations we have used the nominal V magnitude of the three stars. 
We find that for small spots $300.0\;K$ cooler than the photosphere, the correction of the planetary spectrum is, in general, not necessary for all the three analyzed cases. 
For higher temperature contrast ($\Delta T=T_*-T_s=1000.0\;K$), the planetary signal is systematically overestimated if photospheric activity is not properly corrected, especially in the visible band where the spot-star contrast is larger. In order to correct the transit depth, we recover the spectrum of the unspotted photosphere from out-of-transit observations and  compare the difference between the out-of-transit and the in-transit spectrum with the spectrum of the unspotted photosphere (see Eq.~\ref{extraction_planet}). We assume that the distribution of spots on the visible stellar hemisphere does not vary during the transit time scale. 

We have shown that our method can distinguish  planetary atmospheres with and without water traces and that it  is robust with respect to a realistic spot distribution, demonstrating that in the case of a two-spots system we retrieve “effective” spot parameters that allow to extract the planetary spectrum. Moreover the method works better when the stellar parameters are better characterized. This result strongly shows the need to have a reliable and uniform characterization of the ARIEL stellar targets parameters.

Finally, we have estimated the effect of using our method, developed to correct for non-occulted spots, to transits over the spots. The results are given as a function of the spot parameters. For the cases of high S/N ratio (HD 17156 and HAT-P-11 planets) and with a high contrast spot, our method underestimates the planetary signal; the bias is here identified as the median value of the residuals of the retrieved planetary signals. Furthermore, in these cases, the residuals show spurious slopes below $2\mu m$, where the distortions of the stellar spectrum due to the spot increase. However, the underestimate is small, comparable with noise, when the temperature contrasts and the filling factors are small, while for
higher contrast spot the underestimate is significant, but the bump  in the transit light curve is well evident,  can be identified and removed from the data. These effects are indistinct for the planet K2-21 b due to the very low S/N.\\

We plan to develop and expand our method; the next step in this project will be improving and making more detailed the model of stellar activity, by including other photospheric inhomogeneities and limb darkening effects. 

\section*{Acknowledgements}
The authors acknowledge the support of the ARIEL ASI-INAF agreement n.2018-22-HH.0. The authors thank Enzo Pascale and Lorenzo Mugnai from the University La Sapienza of Rome for providing us the ArielRad software simulator. The authors acknowledge the anonymous referee for his/her useful and stimulating comments.

\section*{Data availability}
Data used in this paper are derived from simulations of publicly available codes: \url{https://phoenix.ens-lyon.fr/Grids/BT-Settl/} (\citealt{Baraffe}), \url{https://github.com/ucl-exoplanets/TauREx3_public} (\citealt{TauREx}) and the ArielRad software simulator (\citealt{ArielRad}). The specific simulation can be shared on reasonable request to the corresponding author.

\appendix
\section{Transits over the spots}\label{derivation}
\begin{figure}
    \centering
    \includegraphics[width=\columnwidth]{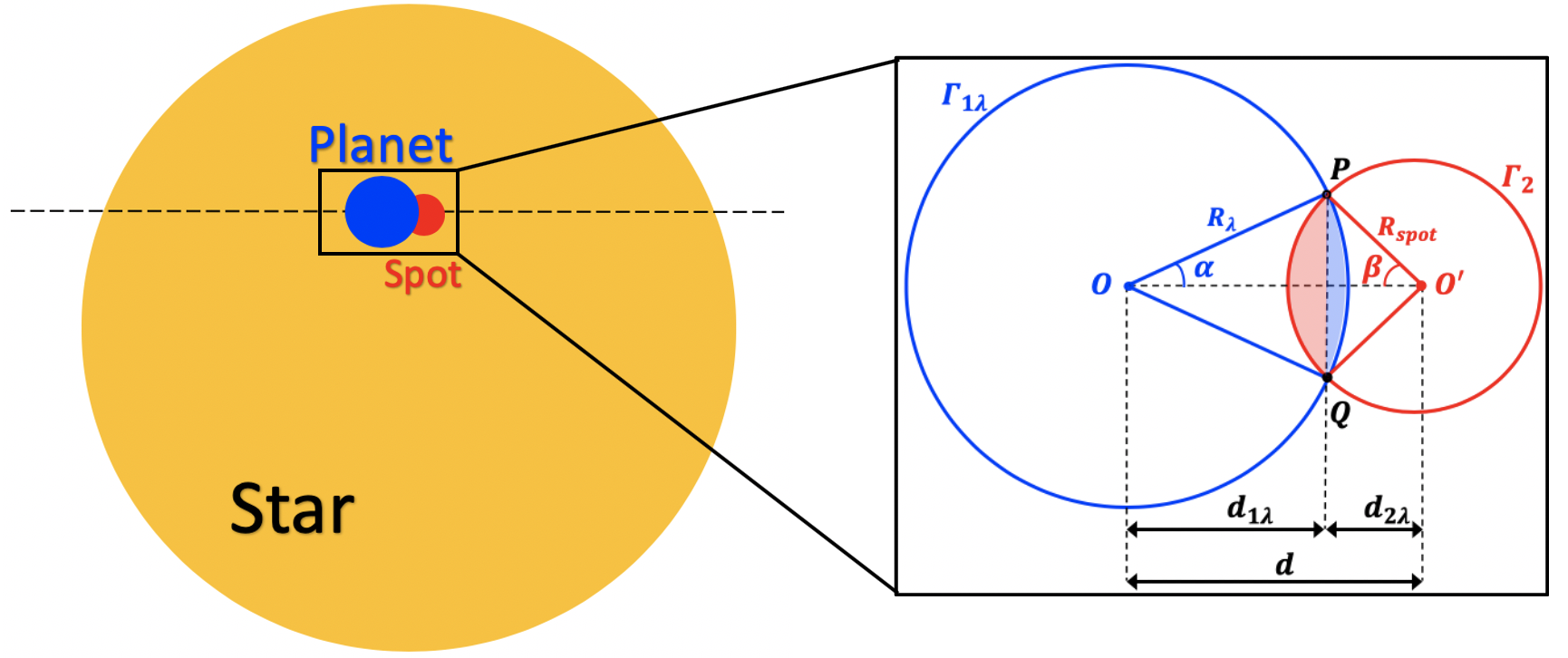}
    \caption{Representative image of the spot crossing. The spot is circular and is in the transit chord. The right box is a zoom on the spot crossing. $\Gamma_{1\lambda}$ and $\Gamma_2$ are the circles representing the projections on the stellar disk of the planet, with radius $R_\lambda$, and of the spot, with radius $R_{spot}$, respectively; $d$ the distance between their centers $O$ and $O'$. The overlapping area between the spot and the planet is represented by the colored region.}
    \label{Intersection}
\end{figure}

In this Appendix we derive Eq.~(\ref{intersection}), which provides the intersection area $g_\lambda \varepsilon_\lambda$ between the spot and the planet at each exposure. Fig.~\ref{Intersection} shows the geometry of the spot crossing, with a zoom on the spot occultation. 
For simplicity, we assume that the spot projection on the stellar disk is a circle with radius $R_{spot}$, while $R_\lambda$ is the planetary radius, whose projection is a circle too. Be $\Gamma_{1\lambda}$ and $\Gamma_2$ the two circles representing the planet and the spot projections respectively, and $d$ the distance between their centers $O$ and $O'$. The simplest cases are:

\begin{itemize}
    \item $R_{spot}+R_\lambda \leq d$: the spot and the planet do not overlap and the intersection area is $\mathcal{A}=0$;
    \item $R_{spot}\leq R_\lambda$ and $d\leq R_\lambda-R_{spot}$: the planet is completely covering the spot, so the intersection area is $\mathcal{A}=\pi R_{spot}^2$;
    \item $R_{spot}>R_\lambda$ and $d\leq R_{spot}-R_\lambda$: the planet is entirely over the spot, so the overlapping area is $\mathcal{A}=\pi R_\lambda^2$.
\end{itemize} 

The most general case is when there is a partial overlap between the spot and the planet ($\left| R_\lambda - R_{spot}\right| < d < R_\lambda + R_{spot}$). We define $P\hat{O}O'=\alpha$ and $P\hat{O'}O=\beta$ (Fig.~\ref{Intersection}); here we will discuss the case where both $\alpha$ and $\beta$ are acute angles. This case occurs as long as:
\begin{equation}
  d \leq \sqrt{\left|R_\lambda^2-R_{spot}^2\right|}
\end{equation}
We will derive Eq.~(\ref{intersection}) in this configuration, but it can be proved by using a similar approach also for the other possible configurations ($\alpha$ acute and $\beta$ obtuse or vice versa).\\
From Fig.~\ref{Intersection}, the intersection area $\mathcal{A}$ can be written as:
\begin{equation}
  \label{n_17}
  \mathcal{A} = \mathcal{A}(PQ_{\Gamma_{1\lambda}}) + \mathcal{A}(PQ_{\Gamma_2})
\end{equation}
where $\mathcal{A}(\cdot)$ indicates the area of the object in brackets, $PQ_{\Gamma_{1\lambda}}$ and $PQ_{\Gamma_2}$ are the two circular segments delimited by the chord PQ (the blue and red region of the zoom in Fig.~\ref{Intersection}). The previous equation may be rewritten as:
\begin{equation}
  \label{n_18}
    \begin{split}
    \mathcal{A} &= \mathcal{A}(OPQ_{c.s.})-\mathcal{A}(OPQ_t) \\
    &+ \mathcal{A}(O'PQ_{c.s.})-A(O'PQ_t)\\
    \end{split}
\end{equation}
where the subscripts $c.s.$ and $t$ stand for “circular sector” and “triangle”, respectively. The areas of the two circular sectors are respectively:
\begin{equation}
  \label{n_19}
  \begin{split}
    \mathcal{A}(OPQ_{c.s.}) &= R_\lambda^2\alpha\\
    \mathcal{A}(O'PQ_{c.s.}) &= R_{spot}^2\beta\\
    \end{split}
\end{equation}
By applying the Pythagorean theorem we derive:
\begin{equation}
  \label{n_21}
  \begin{split}
    \mathcal{A}(OPQ_t) & = d_{1\lambda}\sqrt{R_\lambda^2-d_{1\lambda}^2} \\
    \mathcal{A}(O'PQ_t) & = d_{2\lambda}\sqrt{R_{spot}^2-d_{2\lambda}^2} \\
    \end{split}
\end{equation}
where $d_{1\lambda}$ and $ d_{2\lambda}$ are respectively:

\begin{equation}
    \label{eq_for_d}
\begin{split}
    d_{1\lambda} & =R_\lambda \cos{\alpha}\\
    d_{2\lambda} & =d-d_{1\lambda}\\
\end{split}
\end{equation}
We express the intersection area $\mathcal{A}$ in terms of $R_\lambda$, $R_{spot}$ and $d$.
From Fig.~\ref{Intersection}:
\begin{equation}
  \label{n_1}
  \begin{cases} 
    PQ=2R_\lambda \sin{\alpha} \\ 
    PQ=2R_{spot} \sin{\beta} \\ 
    d=R_\lambda \cos{\alpha}+R_{spot} \cos{\beta}  \\ 
  \end{cases}  
\end{equation}

From the first two relations of Eq.~(\ref{n_1}) and the fundamental trigonometric identity we obtain:

\begin{equation}
  \label{n_2}
    \sin{\alpha}=\frac{R_{spot}}{R_\lambda}\sin{\beta}
\end{equation}

\begin{equation}
  \label{n_3}
     \cos{\alpha}=\sqrt{1-\frac{R_{spot}^2}{R_\lambda^2}\sin^2{\beta}}
\end{equation}
By squaring the third relation of Eq.~(\ref{n_1}):

\begin{equation}
  \label{n_4}
     d^2=R_\lambda^2 \cos^2{\alpha}+R_{spot}^2 \cos^2{\beta}+2R_\lambda R_{spot}\cos{\alpha}\cos{\beta}
\end{equation}
Through Eq.~(\ref{n_2}), Eq.~(\ref{n_3}) and fundamental trigonometric identities we rewrite Eq.~(\ref{n_4}) using just the trigonometric function $\sin{\beta}$:
\begin{equation}
  \label{n_5}
  \begin{split}
     d^2 & =R_\lambda^2 - R_\lambda^2 \sin^2{\alpha}+R_{spot}^2-R_{spot}^2 \sin^2{\beta}\\
     &+2R_\lambda R_{spot}\cos{\alpha}\cos{\beta}\\
     \end{split}
\end{equation}

\begin{equation}
  \label{n_6}
  \begin{split}
     d^2 & =R_\lambda^2 + R_{spot}^2 - 2R_{spot}^2 \sin^2{\beta}\\
     & +2R_\lambda R_{spot}\sqrt{1-\frac{R_{spot}^2}{R_\lambda^2}\sin^2{\beta}}\sqrt{1-\sin^2{\beta}}\\
   \end{split}
\end{equation}

The following steps consist in isolating the term with the radicals in Eq.~(\ref{n_6}), squaring both sides of the equation thus obtained, calculating the products and rearranging in order to derive $\sin{\beta}$ as a function of $R_\lambda$, $R_{spot}$ and $d$:

\begin{equation}
  \label{n_7}
  \begin{split}
     &  \frac{d^2 - R_\lambda^2 - R_{spot}^2}{2R_\lambda R_{spot}}+ \frac{R_{spot}}{R_\lambda}\sin^2{\beta}  =\\ 
     & \sqrt{1-\frac{R_{spot}^2}{R_\lambda^2}\sin^2{\beta}}\sqrt{1-\sin^2{\beta}} \\
    \end{split}
\end{equation}

\begin{equation}
  \label{n_8}
  \begin{split}
     & \left(\frac{d^2 - R_\lambda^2 - R_{spot}^2}{2R_\lambda R_{spot}}\right)^2+ \frac{R_{spot}^2}{R_\lambda^2}\sin^4{\beta}+\frac{d^2 - R_\lambda^2 - R_{spot}^2}{R_\lambda^2}\sin^2{\beta}  =\\
     & 1-\frac{R_\lambda^2+R_{spot}^2}{R_\lambda^2}\sin^2{\beta} + \frac{R_{spot}^2}{R_\lambda^2}\sin^4{\beta}\\
     \end{split}
\end{equation}

\begin{equation}
  \label{n_9}
     \frac{d^2}{R_\lambda^2}\sin^2{\beta}=1-\left(\frac{d^2 - R_\lambda^2 - R_{spot}^2}{2R_\lambda R_{spot}}\right)^2
\end{equation}

\begin{equation}
  \label{n_10}
     \sin^2{\beta}=\frac{-d^4-R_{spot}^4-R_\lambda^4+2d^2R_{spot}^2+2d^2R_\lambda^2+2R_\lambda^2 R_{spot}^2}{4d^2R_{spot}^2}
\end{equation}

\begin{equation}
  \label{n_11}
     \sin{\beta}=\frac{\sqrt{-d^4-R_{spot}^4-R_\lambda^4+2d^2R_{spot}^2+2d^2R_\lambda^2+2R_\lambda^2 R_{spot}^2}}{2d R_{spot}}
\end{equation}
Thus:
\begin{equation}
  \label{n_12}
     \cos^2{\beta}=\frac{d^4+R_{spot}^4+R_\lambda^4+2d^2R_{spot}^2-2d^2R_\lambda^2-2R_\lambda^2 R_{spot}^2}{4d^2R_{spot}^2}
\end{equation}

\begin{equation}
  \label{n_13}
     \cos^2{\beta}=\left(\frac{R_{spot}^2+d^2-R_\lambda^2}{2d R_{spot}}\right)^2
\end{equation}

\begin{equation}
  \label{n_14}
     \cos {\beta}=\frac{R_{spot}^2+d^2-R_\lambda^2}{2d R_{spot}} 
\end{equation}
Substituting Eq.~(\ref{n_10}) and Eq.~(\ref{n_11}) in Eq.~(\ref{n_2}) and Eq.~(\ref{n_3}), similarly we obtain:

\begin{equation}
  \label{n_15}
     \sin{\alpha}=\frac{\sqrt{-d^4-R_\lambda^4-R_{spot}^4+2d^2R_\lambda^2+2d^2R_{spot}^2+2R_\lambda^2R_{spot}^2}}{2d R_\lambda}
\end{equation}

\begin{equation}
  \label{n_16}
     \cos {\alpha}=\frac{R_\lambda^2+d^2-R_{spot}^2}{2R_\lambda d} 
\end{equation}
Finally, we can express the areas $\mathcal{A}(OPQ_{c.s.})$,
$\mathcal{A}(O'PQ_{c.s.})$ in Eq.~(\ref{n_19}) as:

\begin{equation}
  \label{n_22}
  \begin{split}
     A(OPQ_{c.s.}) & = R_\lambda^2\arccos{\left(\frac{R_\lambda^2+d^2-R_{spot}^2}{2dR_\lambda}\right)}\\
     A(O'PQ_{c.s.}) & = R_{spot}^2\arccos{\left(\frac{R_{spot}^2+d^2-R_\lambda^2}{2dR_{spot}}\right)}\\
    \end{split}
\end{equation}
while $ d_{1\lambda}$ and $ d_{2\lambda}$ in Eq.~(\ref{eq_for_d}) become respectively:

\begin{equation}
\begin{split}
    d_{1\lambda} & =\frac{R_\lambda^2+d^2-R_{spot}^2}{2d}\\
    d_{2\lambda} & =\frac{R_{spot}^2+d^2-R_\lambda^2}{2d}\\
\end{split}
\end{equation}

By substituting these terms in Eq.~(\ref{n_18}) and by expressing the intersection area as a fraction of the area of the stellar disk $\pi R_*^2$ we obtain Eq.~(\ref{intersection}).\\

\bsp	
\label{lastpage}

\begin{thebibliography}{}
    
    \bibitem[\protect\citeauthoryear{Agol et al.}{2010}]{Algol} Agol E., Cowan N.~B., Knutson H.~A., Deming D., Steffen J.~H., Henry G.~W., Charbonneau D., 2010, \href{https://ui.adsabs.harvard.edu/abs/2010ApJ...721.1861A} {ApJ, 721, 1861}
    
    \bibitem[\protect\citeauthoryear{Aigrain \& Irwin}{2004}]{Aigrain} Aigrain S., Irwin M., 2004,  \href{https://ui.adsabs.harvard.edu/abs/2004MNRAS.350..331A}{MNRAS, 350, 331}

    \bibitem[\protect\citeauthoryear{Al-Refaie et al.}{2019}]{TauREx} Al-Refaie A.~F., Changeat Q., Waldmann I.~P., Tinetti G., 2019, \href{https://ui.adsabs.harvard.edu/abs/2019arXiv191207759A}{arXiv, p.1912.07759}

    \bibitem[\protect\citeauthoryear{Bakos et al.}{2010}]{Bakos} Bakos G. {\'A}., Torres G., P{\'a}l A., Hartman J., Kov{\'a}cs G., Noyes R.~W., Latham D.~W., et al., 2010, \href{https://ui.adsabs.harvard.edu/abs/2010ApJ...710.1724B}{ApJ, 710, 1724}

    \bibitem[\protect\citeauthoryear{Ballerini et al.}{2012}]{Ballerini} Ballerini P., Micela G., Lanza A.~F., Pagano I., 2012, \href{https://ui.adsabs.harvard.edu/abs/2012A&A...539A.140B} {A\&A, 539, A140} 
    
    \bibitem[\protect\citeauthoryear{Baraffe et al.}{2015}]{Baraffe} Baraffe I., Homeier D., Allard F., Chabrier G., 2015, \href{{https://ui.adsabs.harvard.edu/abs/2015A&A...577A..42B}}{A\&A, 577, A42}
    
    \bibitem[\protect\citeauthoryear{B{\'e}ky et al.}{2014}]{Rotation} B{\'e}ky B., Holman M.~J., Kipping D.~M., Noyes R.~W., 2014, \href{https://ui.adsabs.harvard.edu/abs/2014ApJ...788....1B} {ApJ, 788, 1}
    
    \bibitem[\protect\citeauthoryear{Berta et al.}{2011}]{Berta} Berta Z.~K., Charbonneau D., Bean J., Irwin J., Burke C.~J., D{\'e}sert J.-M., Nutzman P., et al., 2011, \href {https://ui.adsabs.harvard.edu/abs/2011ApJ...736...12B}{ApJ, 736, 12}

    \bibitem[\protect\citeauthoryear{Bonomo \& Lanza}{2008}]{Bonomo_2008} Bonomo A.~S., Lanza A.~F., 2008, \href{https://ui.adsabs.harvard.edu/abs/2008A&A...482..341B} {A\&A, 482, 341}

    \bibitem[\protect\citeauthoryear{Bonomo et al.}{2009}]{Bonomo_2009} Bonomo A.~S., Aigrain S., Bord{\'e} P., Lanza A.~F., 2009, \href{https://ui.adsabs.harvard.edu/abs/2009A&A...495..647B} {A\&A, 495, 647}
    
    \bibitem[\protect\citeauthoryear{Czesla et al.}{2009}]{Czesla} Czesla S., Huber K.~F., Wolter U., Schr{\"o}ter S., Schmitt J.~H.~M.~M., 2009, \href {https://ui.adsabs.harvard.edu/abs/2009A&A...505.1277C} {A\&A, 505, 1277} 

    \bibitem[\protect\citeauthoryear{D{\'e}sert et al.}{2011}]{Desert} D{\'e}sert J.-M., Sing D., Vidal-Madjar A., H{\'e}brard G., Ehrenreich D., Lecavelier Des Etangs A., Parmentier V., et al., 2011, \href{https://ui.adsabs.harvard.edu/abs/2011A&A...526A..12D} {A\&A, 526, A12}

    \bibitem[\protect\citeauthoryear{Edwards et al.}{2019}]{Edwards} Edwards B., Mugnai L., Tinetti G., Pascale E., Sarkar S., 2019, \href{https://ui.adsabs.harvard.edu/abs/2019AJ....157..242E} {AJ, 157, 242}

    \bibitem[\protect\citeauthoryear{Kreidberg et al.}{2014}]{Kreidberg_2014} Kreidberg L., Bean J.~L., D{\'e}sert J.-M., Benneke B., Deming D., Stevenson K.~B., Seager S., et al., 2014, \href{https://ui.adsabs.harvard.edu/abs/2014Natur.505...69K} {Natur, 505, 69}

    \bibitem[\protect\citeauthoryear{McCullough et al.}{2014}]{McCullough} McCullough P.~R., Crouzet N., Deming D., Madhusudhan N., 2014, \href{https://ui.adsabs.harvard.edu/abs/2014ApJ...791...55M} {ApJ, 791, 55} 
    
    \bibitem[\protect\citeauthoryear{Micela}{2015}]{Micela} Micela G., 2015, \href{https://ui.adsabs.harvard.edu/abs/2015ExA....40..723M} {ExA, 40, 723} 

    \bibitem[\protect\citeauthoryear{Morris et al.}{2017}]{Morris} Morris B.~M., Hebb L., Davenport J.~R.~A., Rohn G., Hawley S.~L., 2017, \href {https://ui.adsabs.harvard.edu/abs/2017ApJ...846...99M} {ApJ, 846, 99}

    \bibitem[\protect\citeauthoryear{Moutou et al.}{2005}]{Moutou} Moutou C., Pont F., Barge P., Aigrain S., Auvergne M., Blouin D., Cautain R., et al., 2005, \href{https://ui.adsabs.harvard.edu/abs/2005A&A...437..355M} {A\&A, 437, 355} 
    
    \bibitem[\protect\citeauthoryear{Mugnai et al.}{2020}]{ArielRad} Mugnai L.~V., Pascale E., Edwards B., Papageorgiou A., Sarkar S., 2020, \href{https://ui.adsabs.harvard.edu/abs/2020ExA....50..303M}{ExA, 50, 303}

    \bibitem[\protect\citeauthoryear{Nelder \& Mead}{Nelder \& Mead}{1965}]{Nelder} Nelder J.~A.,  Mead R.,  1965, \href{https://doi.org/10.1093/comjnl/7.4.308}{The Computer Journal, 7, 308–313}

    \bibitem[\protect\citeauthoryear{O'Neal, Neff, \& Saar}{1998}]{O_Neal_1998} O'Neal D., Neff J.~E., Saar S.~H., 1998, \href {https://ui.adsabs.harvard.edu/abs/1998ApJ...507..919O}{ApJ, 507, 919} 
    
    \bibitem[\protect\citeauthoryear{{Pearson} et~al.,}{{Pearson}
      et~al.}{2020}]{Data_level}
    {Pearson} C.,  et~al., 2020, Paper submitted to ExA
    
    \bibitem[\protect\citeauthoryear{Pont et al.}{2013}]{Pont} Pont F., Sing D.~K., Gibson N.~P., Aigrain S., Henry G., Husnoo N., 2013, \href {https://ui.adsabs.harvard.edu/abs/2013MNRAS.432.2917P}{MNRAS, 432, 2917}

    \bibitem[\protect\citeauthoryear{Salz et al.}{2018}]{Salz_2018} Salz M., Czesla S., Schneider P.~C., Nagel E., Schmitt J.~H.~M.~M., Nortmann L., Alonso-Floriano F.~J., et al., 2018, \href{https://ui.adsabs.harvard.edu/abs/2018A&A...620A..97S}{A\&A, 620, A97} 
    
    \bibitem[\protect\citeauthoryear{Scandariato \& Micela}{2015}]{Scandariato2015} Scandariato G., Micela G., 2015, \href{https://ui.adsabs.harvard.edu/abs/2015ExA....40..711S} {ExA, 40, 711} 

    \bibitem[\protect\citeauthoryear{Seager \& Mall{\'e}n-Ornelas}{2003}]{Seager} Seager S., Mall{\'e}n-Ornelas G., 2003, \href {https://ui.adsabs.harvard.edu/abs/2003ApJ...585.1038S}{ApJ, 585, 1038} 
    
    \bibitem[\protect\citeauthoryear{Silva-Valio et al.}{2010}]{CoRoT-2} Silva-Valio A., Lanza A.~F., Alonso R., Barge P., 2010, \href {https://ui.adsabs.harvard.edu/abs/2010A&A...510A..25S}{A\&A, 510, A25} 

    \bibitem[\protect\citeauthoryear{Sing}{2018}]{sing_2018} Sing D.~K., 2018, \href{https://ui.adsabs.harvard.edu/abs/2018arXiv180407357S} {arXiv, p. 1804.07357}

    \bibitem[\protect\citeauthoryear{Sing et al.}{2009}]{Sing_2009} Sing D.~K., D{\'e}sert J.-M., Lecavelier Des Etangs A., Ballester G.~E., Vidal-Madjar A., Parmentier V., Hebrard G., et al., 2009, \href{https://ui.adsabs.harvard.edu/abs/2009A&A...505..891S} {A\&A, 505, 891}


    \bibitem[\protect\citeauthoryear{Tabernero et al.}{2019}]{Example} Tabernero H.~M., Marfil E., Montes D., Gonz{\'a}lez Hern{\'a}ndez J.~I., 2019,  
    \href{https://ui.adsabs.harvard.edu/abs/2019A&A...628A.131T} {A\&A, 628, A131}
    
    \bibitem[\protect\citeauthoryear{Tinetti, Encrenaz, \& Coustenis}{2013}]{description_observation} Tinetti G., Encrenaz T., Coustenis A., 2013, \href{https://ui.adsabs.harvard.edu/abs/2013A&ARv..21...63T} {A\&ARv, 21, 63}
    
    \bibitem[\protect\citeauthoryear{Tinetti et al.}{2016}]{ARIEL} Tinetti G., Drossart P., Eccleston P., Hartogh P., Heske A., Leconte J., Micela G., et al., 2016, \href{https://ui.adsabs.harvard.edu/abs/2016SPIE.9904E..1XT}{SPIE, 9904, 99041X}
    
    \bibitem[\protect\citeauthoryear{Tinetti et al.}{2018}]{Tinetti_2018} Tinetti G., Drossart P., Eccleston P., Hartogh P., Heske A., Leconte J., Micela G., et al., 2018, \href{https://ui.adsabs.harvard.edu/abs/2018ExA....46..135T}{ExA, 46, 135} 

    \bibitem[\protect\citeauthoryear{Virtanen et al.}{2020}]{scipy} Virtanen P., Gommers R., Oliphant T.E., Haberland M., Reddy T., Cournapeau D.,
    Burovskin E., et al., 2020
    \href{https://rdcu.be/b08Wh}{Nature Methods, 17, 261-272}
    
    \bibitem[\protect\citeauthoryear{Zellem et al.}{2017}]{Zellem} Zellem R.~T., Swain M.~R., Roudier G., Shkolnik E.~L., Creech-Eakman M.~J., Ciardi D.~R., Line M.~R., et al., 2017,
    \href{https://ui.adsabs.harvard.edu/abs/2017ApJ...844...27Z} {ApJ, 844, 27}
\end{thebibliography}
\end{document}